\documentclass[a4paper,11pt]{article}
\pdfoutput=1 

\usepackage{jcappub} 
\usepackage[T1]{fontenc}
\usepackage{amssymb}
\usepackage{amsmath}
\usepackage{epsfig}
\usepackage{graphicx}
\usepackage{natbib}

\renewcommand{\vec}[1]{\mathbf{#1}}

\newcommand{\vk}{\vec{k}}

\newcommand*{\df}  {\delta}

\newcommand*{\non}  {\nonumber}
\newcommand*{\lb}  {\left(}
\newcommand*{\rb}  {\right)}

\newcommand*{\la}  {\left\langle}
\newcommand*{\ra}  {\right\rangle}

\newcommand{\ba}{\[\begin{aligned}}
\newcommand{\ea}{\end{aligned}\]}

\newcommand{\eq}[1]{\begin{align}#1\end{align}}
\newcommand{\eeq}[1]{\begin{equation}#1\end{equation}}

\makeatletter
\DeclareFontFamily{OMX}{MnSymbolE}{}
\DeclareSymbolFont{MnLargeSymbols}{OMX}{MnSymbolE}{m}{n}
\SetSymbolFont{MnLargeSymbols}{bold}{OMX}{MnSymbolE}{b}{n}
\DeclareFontShape{OMX}{MnSymbolE}{m}{n}{
    <-6>  MnSymbolE5
   <6-7>  MnSymbolE6
   <7-8>  MnSymbolE7
   <8-9>  MnSymbolE8
   <9-10> MnSymbolE9
  <10-12> MnSymbolE10
  <12->   MnSymbolE12
}{}
\DeclareFontShape{OMX}{MnSymbolE}{b}{n}{
    <-6>  MnSymbolE-Bold5
   <6-7>  MnSymbolE-Bold6
   <7-8>  MnSymbolE-Bold7
   <8-9>  MnSymbolE-Bold8
   <9-10> MnSymbolE-Bold9
  <10-12> MnSymbolE-Bold10
  <12->   MnSymbolE-Bold12
}{}

\let\llangle\@undefined
\let\rrangle\@undefined
\DeclareMathDelimiter{\llangle}{\mathopen}%
                     {MnLargeSymbols}{'164}{MnLargeSymbols}{'164}
\DeclareMathDelimiter{\rrangle}{\mathclose}%
                     {MnLargeSymbols}{'171}{MnLargeSymbols}{'171}
\makeatother

\newcommand{\at}{\makeatletter @\makeatother}

\newcommand{\be}{\begin{equation}}
\newcommand{\ee}{\end{equation}}
\newcommand{\bea}{\begin{eqnarray}}
\newcommand{\eea}{\end{eqnarray}}

\author[a,b]{Zvonimir Vlah}
\author[c,d,e]{Uro\v s Seljak}
\author[e]{Man Yat Chu}
\author[e]{Yu Feng}

\affiliation[a]{Stanford Institute for Theoretical Physics and Department of Physics, Stanford University, Stanford, CA 94306, USA} 
\affiliation[b]{Kavli Institute for Particle Astrophysics and Cosmology, SLAC and Stanford University, Menlo Park, CA 94025, USA}
\affiliation[c]{Physics, Astronomy Department, University of California, Berkeley, CA 94720, USA} 
\affiliation[d]{Lawrence Berkeley National Laboratory, Berkeley, CA 94720, USA}
\affiliation[e]{Berkeley Center for Cosmological Physics, University of California, Berkeley, CA 94720, USA} 

\emailAdd{zvlah\at stanford.edu}
\emailAdd{useljak\at berkeley.edu}
\emailAdd{jack\_chu\at berkeley.edu}
\emailAdd{yfeng1\at berkeley.edu}

\title{Perturbation theory, effective field theory, and oscillations in the power spectrum}


\arxivnumber{1509.02120}

\abstract{
We explore the relationship between the nonlinear matter power spectrum 
and the various Lagrangian and Standard Perturbation Theories (LPT and SPT).
We first look at it in the context of one dimensional (1-d) dynamics, where 1LPT 
is exact at the perturbative level and one can exactly resum the SPT series into 
the 1LPT power spectrum. Shell crossings lead to non-perturbative effects, and 
the PT ignorance can be quantified in terms of their ratio, which is also the transfer 
function squared in the absence of stochasticity. 
At the order of PT we work, this parametrization is equivalent to the results of 
effective field theory (EFT), and can thus be expanded in terms of 
the same parameters.
We find that its radius of convergence 
is larger than the SPT loop expansion. The same EFT parametrization applies to 
all SPT loop terms and if stochasticity can be ignored, to all N-point correlators. 
In 3-d, the LPT structure is 
considerably more complicated, and we find that LPT models with parametrization motivated 
by the EFT exhibit running with $k$ and that SPT is generally a better choice. 
Since these transfer function expansions contain free parameters that change with cosmological 
model their usefulness for broadband power is unclear. For this reason
we test the predictions of these models on baryonic 
acoustic oscillations (BAO) and other primordial oscillations, including string 
monodromy models,
for which we ran a series of simulations with and without 
oscillations. Most models are successful in predicting oscillations beyond their 
corresponding PT versions, confirming the basic validity of the model. 
We show that if primordial oscillations are localized to a scale $q$, the wiggles 
in power spectrum are approximately suppressed as $\exp[-k^2\Sigma^2(q)/2]$, 
where $\Sigma(q)$ is rms displacement of particles separated by $q$, which saturates 
on large scales, and decreases as $q$ is reduced. 
No oscillatory features survive past $k \sim 0.5$h/Mpc at $z=0$. 
}

\begin{document}
\maketitle
\flushbottom


\section{Introduction}

Effective Field Theory (EFT) approach to large scale structure (LSS) \cite{2012JCAP...07..051B,
2012JHEP...09..082C, 2013JCAP...08..037P, 2015JCAP...02..013S, 2014JCAP...05..022P, 
2014JCAP...07..057C, 2015arXiv150705326F} 
has lately received a lot of attention as a way to extend the validity of cosmological 
perturbation theory (PT) (see e.g. \cite{2002PhR...367....1B, 2013MNRAS.429.1674C, 2014JCAP...01..010B}), 
at a cost of introducing free 
parameters (e.g. \cite{2012JHEP...09..082C, 2014JCAP...05..022P,2014JCAP...07..057C}). 
These parameters must obey some symmetry requirements. 
For example, any local nonlinear scrambling of matter must obey mass and 
momentum conservation \cite{1980lssu.book.....P}, and one can show that at 
lowest $k$ (where $k$ is the wavevector amplitude of the Fourier modes), 
the leading order effects scale as $\alpha k^2P_L(k)$, where $P_L(k)$ is the 
linear matter power spectrum. In the halo model \cite{2000MNRAS.318..203S,2001ApJ...546...20S,
2002PhR...372....1C, 2011A&A...527A..87V} one can assemble the local 
scrambling of dark matter into dark matter halos, and one can perform a Taylor series 
expansion of the halo profile to extend this into a series of even powers of $k$ 
\cite{2014MNRAS.445.3382M}. There is a second term that in EFT language is called the stochastic 
or mode coupling term, for which mass and momentum conservation require to initially 
scale as $k^4$. This term is often ignored in EFT calculations, but it 
should be the dominant term on small scales. Since PT calculations already 
enforce the conservation laws such terms can be applied to any PT scheme 
as a way to correct for whatever is missing in PT calculations. So EFT can be 
viewed simply as a parametrization of the ignorance of a given PT model, 
absorbing any discrepancy between the true solution and the PT solution, 
and parametrizing it in terms of a simple parameter expansion. 

The usefulness of this approach then depends on the convergence radius of 
this expansion, or more simply, the range of $k$ over which the lowest order 
EFT term(s) restore the exact solution. The answer will depend on the specific 
PT model implementation: one can apply PT ignorance to linear theory, for example, 
but that will not be very useful and will lead to a strong running of the EFT 
parameter with scale, as well as a large stochasticity \cite{2012JCAP12011T, 2015arXiv150702255B}. 
This is because the lowest order EFT correction is of the same order as PT at the 1-loop order, 
so it makes sense to add lowest order EFT term to 1-loop PT. 
Moreover, we will argue that the 
convergence radius of EFT expansion should be a bit larger than that of loop 
expansion.  While this discussion is general, a very relevant question is which 
PT to use. In literature we have PT approaches both in Eulerian space (SPT) 
(see e.g. \cite{2002PhR...367....1B, 2014JCAP...07..057C, 2014JCAP...01..010B,2015arXiv150706665B},
and in Lagrangian space (LPT) 
(see e.g. \cite{2013MNRAS.429.1674C, 2014JCAP...05..022P}), 
and at several different orders. These 
give very different predictions for the power spectrum, and it is unclear which 
is more successful. The purpose of this paper is twofold. One is simply to test 
the various PT models, parametrize their ignorance against the true answer 
in terms of EFT parameters, and study their scale dependence. The less scale 
dependence there is, the more useful the expansion. Our goal is to test several 
PT models, including some introduced here for the first time. 

Moreover, scale dependence alone is not the only criterion, as it could be a coincidence 
that the EFT parameter is roughly constant over a certain range of $k$. 
Our second purpose is to test the validity of EFT+PT approaches by applying 
it to modeling of oscillatory features in linear power spectrum. 
Baryonic acoustic oscillations (BAO) are a prime example of usefulness 
of EFT approach: BAO appear at low $k$, where we would expect EFT 
corrections to be valid. Since EFT corrections at the lowest order scale with 
$P_L(k)$, and $P_L(k)$ contains BAO, one would expect EFT correction to 
carry the signature of BAO. 
EFT correction is a small effect on top of a small BAO effect, 
so we would not expect it to be visible in simulations at low $k$, where sampling 
variance errors dominate. For this reason we ran a series of simulations with and 
without oscillations, but with the same initial conditions (as described in \cite{2015PhRvD..91b3508V}), 
so that the sampling variance errors cancel \cite{2009JCAP...10..007M}. 

In this paper, we adapt the simple parametrization of the dark matter overdensity via 
the density transfer function (see e.g.  \cite{2012JCAP12011T}). In this approach, 
EFT parameters are obtained by expanding the transfer function in even powers of k. 
This equivalence of EFT and transfer function expansions holds 
at the lowest order in PT (see e.g. \cite{2014JCAP...07..057C}), but not beyond that. 
This means that the leading coefficient of transfer function expansion is $k^2$, but beyond that 
one can have an arbitrary Taylor expansion. Nevertheless, for simplicity we will try even powers of $k$
as our expansion basis. Alternatively, one could also try different transfer functions at 
higher orders in density (this was recently 
indicated in \cite{2015arXiv150702255B}). 
With all this in mind, for simplicity 
we will refer to all such expansions as the EFT expansions, and parameters we will call 
EFT parameters.

The outline of the paper is as follows. The relationship between LPT, SPT, 
and EFT is particularly simple in 1-d dark matter dynamics and in section 
\ref{1d} we first analyze this example using results from \cite{2015arXiv150207389M}. 
We then move to 3-d analysis in section \ref{3d}, extracting EFT parameters 
for various PT models. In section \ref{bao} we apply these to the modeling of 
BAO features and other oscillations, and in section \ref{pk} we discuss the 
general lessons for the modeling of power spectrum. We present the conclusions 
in section \ref{conc}. In appendix \ref{app:nowiggle} we present the details of the 
construction of no-wiggle power spectrum.

\section{Investigation of the 1-d example}
\label{1d}

To introduce the different schemes and their relation we will first look at the 1-d 
example. It is useful to look at the PT expansions in the context of 1-d dynamics, 
where 1LPT (Zeldovich) solution is exact at the perturbative level, and can be 
shown to be identical to SPT in the infinite loop limit \cite{2015arXiv150207389M}, 
\be
P_{{\rm 1LPT}}(k,z)=\sum_{i=0}^{\infty} D_+^{2(i+1)}(z)P_{{\rm SPT},i-{\rm loop}}(k,z=0) , 
\ee
where $P_{{\rm SPT},i-{\rm loop}}$ is the $i$-loop SPT term with 0-th loop given by $P_L(k)$, 
the linear power spectrum. These terms are multiplied with the appropriate growth 
rate $D_+(z)$, which we will for convenience normalize to $D_+(z=0)=1$, so they 
can be dropped as long as we work at $z=0$. From results of \cite{2015arXiv150207389M} 
we estimate the radius of convergence of SPT to be around (0.2-0.3)h/Mpc for 
$\Lambda$CDM like linear power spectra (in terms of power per mode): the scale 
at which the 10 loop SPT is accurate is $k \sim 0.2$h/Mpc, and SPT series becomes 
highly oscillatory for $k > 0.2$h/Mpc. 

1LPT solution is however not the true solution to the dark matter $P_{\rm dm}(k)$, 
a consequence of shell crossings, at which 1LPT sheets continue to stream through 
the shell crossings unperturbed, while in the actual dark matter dynamics they stick 
together inside the high density sheets, the equivalent of halos in 3-d. In 1-d the 
1LPT solution is exact up to shell crossings, and it agrees with N-body simulations 
away from high density regions: the only difference is in the regions of shell crossings, 
which are broader in 1LPT solution and typically have double peaks: 1LPT solution 
artificially spreads out the true density field. In the halo model language 
\cite{2000MNRAS.318..203S,2001ApJ...546...20S, 2002PhR...372....1C} the 
leading term correction at low $k$ comes from applying this smearing to the true 
density profiles of dark matter, which gives rise to the even powers of $k$ series 
corrections multiplying the linear power spectrum $P_L(k)$, and the same is found 
in the context of EFT (e.g. \cite{2012JHEP...09..082C, 2014JCAP...05..022P}). 
The leading order correction is thus the 2-halo (i.e. EFT like) 
$\alpha k^2P_L$, with a positive sign to compensate for excessive smearing of 
1LPT displacements. Here $\alpha^{1/2} $ corresponds to a typical scale 
of the streaming beyond the shell crossings, which can be several Mpc/h at $z=0$. 
These stream crossing induced nonlinear corrections to $P_{1LPT}(k)$ are very large, 
10\% at $k=0.07$h/Mpc and growing to a factor of 2 at $k=0.3$h/Mpc at $z=0$. 
In 1-d it is clear that PT cannot address these stream crossings. Note however that 
all the deviations from 1LPT are within a few Mpc/h. Hence, while the nonlinear 
effects are large in the power spectrum down to very low $k$, the correlation 
function can still be very close to 1LPT on scales larger than a few Mpc 
(away from BAO peak) \cite{2015arXiv150207389M}. 

Even in the presence of stream crossings the 1LPT field is well correlated 
with the dark matter field, a consequence of the fact that it spreads the high
density peaks into a double peaked structure with an approximately constant 
radius independent of the position (and independent of the collapsed mass in 
the sheets). This can be quantified by introducing the transfer function 
(see e.g. \cite{2012JCAP12011T})
\be
\tilde{T}_{{\rm 1LPT}}(k)={\langle \delta_{{\rm 1LPT}} \delta_{\rm dm} \rangle 
\over \langle \delta_{{\rm 1LPT}} \delta_{{\rm 1LPT}} \rangle },
\ee
where $\delta_{\rm 1LPT}$ and $\delta_{\rm dm}$ are the 1LPT and dark matter
density perturbations in Fourier space, respectively. 
We can also introduce the cross-correlation coefficient 
\be
r^2_{{\rm 1LPT}}={\langle \delta_{{\rm 1LPT}} \delta_{\rm dm} \rangle^2 \over \langle 
\delta_{{\rm 1LPT}} \delta_{{\rm 1LPT}} \rangle \langle \delta_{\rm dm} \delta_{\rm dm}\rangle}. 
\ee
It is expected to approach unity at low $k$ and 
it has been shown that  
up to $k=0.2$/Mpc, the stochasticity $1-r_{\rm 1LPT}^2$ is below 1\% in 1-d 
\cite{2015arXiv150207389M}. 
We will denote auto-power spectrum as $ (2\pi)^3 \delta^D(\vk+\vk ') P_{X}(k)
=\langle \delta_{X}(\vk) \delta_{X}(\vk ') \rangle$, where $X$ stands for dm, 
1LPT etc. and $\delta^D$ is Dirac delta function. The corresponding stochastic 
power $P_J(k)$ is defined as
\be
P_J(k)=\big[ 1-r^2_{{\rm 1LPT}}(k) \big]P_{\rm dm}(k).
\ee
The transfer function has to obey some symmetry properties, and in 
particular has to start as $k^2$ \cite{1980lssu.book.....P}. 
We will expand it into a general function of even powers of $k$. 
\label{alphak1d}
If $r_{\rm 1LPT}(k)<1$ then we need a separate function $r_{\rm 1LPT}(k)$ to 
fully describe the dark matter power spectrum given $P_{\rm 1LPT}(k)$, 
and if the goal is to specify the non-perturbative effects on the power 
spectrum then it is simpler to define, 
\be
{P_{{\rm dm}}(k) \over P_{{\rm 1LPT}}(k)} \equiv 
T_{{\rm 1LPT}}^2(k)= 1+\alpha_{{\rm 1LPT}}(k)k^2\equiv \left(1+\sum_{i=1}^{\infty} \alpha_{i,{\rm 1LPT}}k^{2i}\right).
\label{alphak1d2}
\ee
Note that $T(k)$ includes stochasticity and there is no guarantee that it can 
be expanded in terms of even powers of $k$, although we expect that at low 
$k$ the leading term is $\alpha_{\rm 1,1LPT}k^2$. 

\begin{figure*}[tb]
\centering
\includegraphics[scale=0.60 ]{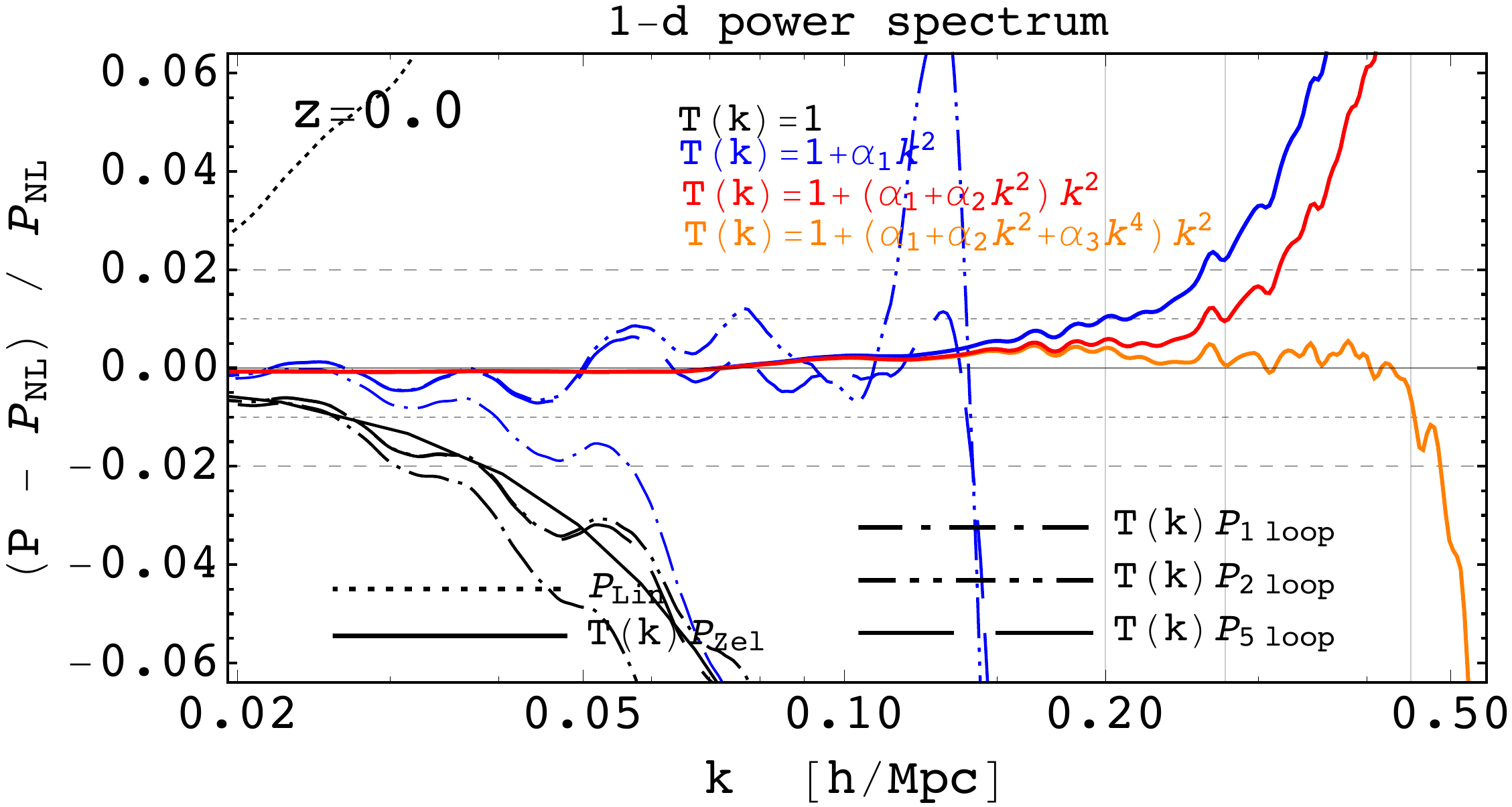}
\caption{
Error of various models of 1-d power spectrum shown relative to the nonlinear simulations results. 
Show are PT results in black: 1LPT/Zeldovich (dashed line), 1-loop SPT 
(dot-dashed line), 2-loop SPT (double dot-dashed line), 5-loop SPT (long-dashed line) 
and linear theory result (dotted line).
In addition we apply the transfer functions to these results giving us EFT+SPT 
and EFT+1LPT models in 1-d. Results for three different transfer functions are shown:
going up to $\alpha_1$ (in blue), $\alpha_2$ (in red) and $\alpha_3$ (in orange)
in expansion given by Eq. \eqref{alphak1d2}. 
Thin grey horizontal dotted and dashed lines represent respectively 1\% and 2\% errors. 
Thin grey vertical solid lines represent maximal $k$ values up to where EFT+1LPT models
acheve 1\% errors. Results are shown at redshift $z=0$.
}
\label{fig1}
\end{figure*}

We have fitted this expansion to the numerical results for $P_{{\rm dm}}(k)$ 
and $P_{\rm 1LPT}(k)$ given in \cite{2015arXiv150207389M}. 
This gives the values of the first three 
coefficients $\alpha_{1,{\rm 1LPT}}=14{\rm (Mpc/h)^2}$, $\alpha_{2,{\rm 1LPT}}=
-4{\rm (Mpc/h)^4}$ and $\alpha_{3,{\rm 1LPT}}=-40{\rm (Mpc/h)^6}$.
This is shown in top of figure \ref{fig1} and the first coefficient is a good fit 
at 1\% level up to $k \sim 0.2$h/Mpc, while with three coefficients the fit 
is good to $k \sim 0.4$h/Mpc. Note that we do not see the value 
$\alpha_{1,1LPT}$ to approach a constant at low $k$: there is no sampling 
variance scatter, but there could be numerical issues with the simulations 
that prevent us from extracting the true value at very low $k$. 
We see the same issue in 3-d (recently also shown in \cite{2015arXiv150702255B}), 
and similar results were found for the displacement analysis of 
\cite{2015arXiv150507098B}. At very low $k$ the nonlinear effects 
are really small, and these issues are unlikely to be relevant for any 
observations, since sampling variance errors are large on large scales.   

From figure \ref{fig1} we see that we need just one EFT parameter even when go to 5th order in SPT 
(and beyond). 
This suggests that the radius of convergence in 
$k$ for SPT expansion is smaller than for EFT expansion. Physically this makes 
sense: SPT breaks down for $\delta_L \sim 1$, and after that the collapse and 
shell crossings occur (the same happens in 3-d, where one usually defines 
$\delta_c \sim 1.68$ as the linear density at collapse). There is some justification 
for the counting of EFT orders to be the same as the SPT loop orders, but the 
two can be different, and one expects EFT expansion to need fewer 
terms for a given SPT order. Finally, note that the stochastic term becomes of 
order 1\% around $k \sim 0.2$/Mpc at $z=0$ and rapidly increases for higher $k$. 
In this regime the expansion in terms of even powers of $k$ in 
equation \ref{alphak1d} is not well justified, 
and instead one needs all powers of $k$, as in any Taylor expansion. 

To see the interplay between the SPT and EFT expansions, 
we can assume $\alpha_{1,{\rm 1LPT}}$ applied to each SPT term and to full 1LPT,  
or we order the terms assuming even powers of $k$ of EFT expansion 
corresponds to a given SPT loop order. We can thus write, assuming $r_{{\rm 1LPT}}=1$, 
\begin{align}
P_{\rm dm}(k)
=\left[1+\alpha_{\rm 1LPT}(k)k^2\right]\sum_{i=0}^{\infty} P_{{\rm SPT,}i-{\rm loop}}(k)
=\sum_{i=0}^{\infty} \sum_{j=0}^i P_{{\rm SPT},(i-j)-{\rm loop}}\alpha_{j,{\rm 1LPT}}k^{2j}
\end{align}
The lowest orders (up to the 2-loop) are, 
\begin{align}
P_{\rm dm}(k)=P_{{\rm L}}(k)
&+P_{{\rm SPT,1-loop}}(k)+\alpha_{1,{\rm 1LPT}}k^2P_{{\rm L}}(k) \nonumber \\
&+P_{{\rm SPT,2-loop}}(k)
  +\alpha_{1,{\rm 1LPT}}k^2P_{{\rm SPT,1-loop}}(k)+\alpha_{2,{\rm 1LPT}}k^4P_{{\rm L}}(k)+\ldots
\end{align}
At 1-loop SPT order the leading correction is $P_{{\rm L}}(k)\alpha_{1,{\rm 1LPT}}k^2$, 
same as in the standard EFT approach \cite{2012JHEP...09..082C}. At 2-loop 
order we pick up two EFT terms in addition to the 2-loop SPT term, 
$\alpha_{1,{\rm 1LPT}} k^2 P_{{\rm SPT,1-loop}}(k) $ and $\alpha_{2,{\rm 1LPT}}k^4P_{{\rm L}}(k)$. 
If the radius of convergence for EFT is  larger than for SPT then we may not need 
the latter term. Note that 2-loop SPT term vanishes at low $k$ in 1-d relative to 
1-loop SPT: this is no longer the case in 3-d, as discussed in next section. 

We use the coefficients derived above and SPT terms from \cite{2015arXiv150207389M}
to plot the error of the different expansions relative to the full solution. 
The results are shown in figure \ref{fig1}. 
Without any EFT parameters the 1LPT (and 1-loop SPT) solution has 
1\% accuracy only to $k=0.03$h/Mpc. With 1 EFT parameter applied to full 
1LPT (i.e. infinite loop SPT) we find 1\% accuracy to $k \sim 0.2$h/Mpc (solid blue line 
in figure \ref{fig1}), and with 3 EFT parameters to $k \sim 0.45$h/Mpc (solid orange line 
in figure \ref{fig1}). In contrast, 1-loop SPT applied to full $\alpha_{1,{\rm 1LPT}}(k)$ 
(i.e. infinite order EFT) is 1\% accurate to $k \sim 0.04$h/Mpc, and 2-loop SPT to 
$k \sim 0.1$h/Mpc. There is no improvement in adding additional EFT parameters 
up to this order. This confirms that the radius of convergence of transfer function 
(EFT) expansion is larger than SPT radius of convergence. Beyond 2-loop the 
improvements in SPT loops are more modest: 5 loop SPT only extends the 
agreement by 10\% in $k$, mostly because 5 loop SPT does not improve much 
the agreement with 1LPT. Overall the EFT gains in combination with SPT are most 
successful at very low $k$, where one expects one EFT term $\alpha_{1,{\rm 1LPT}}$ 
to be sufficient and SPT loop terms are rapidly converging to 1LPT.

There are several lessons of 1-d example worth emphasizing. As shown in \cite{2015arXiv150207389M}
in 1-d case 1LPT is exact solution at the perturbative level since the equation for the displacement field is linear, 
but shell crossings invalidate this solution: hence PT can never fully describe completely the dynamics of dark matter. 
The SPT series can be resummed into 1LPT solution, so 1LPT is superior to 
SPT at any given loop order. One can define a concept of transfer functions 
$T^2_{{\rm 1LPT}}(k)$ that is defined as a ratio of dark matter to 1LPT power spectrum, 
that contains all the information on the power spectrum effects beyond PT. 
One can expand $T_{{\rm 1LPT}}^2(k)$ into a series of powers of $k$ (at low $k$ only even powers contribute) , with 
coefficients that act as EFT parameters. This EFT expansion resummed and 
applied to 1LPT then gives the full dark matter power spectrum. There are thus 
two expansions, one in terms of EFT parameters multiplying even powers of $k$, 
and the second expansion is related to the loop orders of SPT, and the latter has a shorter radius 
of convergence. The same EFT expansion of the transfer function derived from 1LPT 
also applies to each of the SPT loop terms: there is only one EFT expansion in 1-d. 
One can order the two expansions in terms of EFT+SPT order. At 1-loop order 
one finds the usual EFT expression, with the first EFT term ($k^2$ term) 
multiplying $P_{\rm L}(k)$, while at 2-loop order the first EFT term also multiplies 1-loop 
SPT term. Finally, we note that 1-loop SPT+EFT does not extend the range of 
1-loop SPT significantly, only by 30\%, while 2-loop SPT shows a considerable 
improvement, by a factor of 3 in scale. A third scale one can define is the scale 
where stochastic terms become important, and where 1LPT no longer correlates 
well with dark matter. There is no obvious advantage in separating the terms into 
the part that correlates with 1LPT and the part that does not, if one is only 
interested in the power spectrum. This changes if one also includes higher order 
correlations, as discussed below. Many of these features translate into 3-d as 
well, but there are additional complications in 3-d that are described in the next 
section. 

\section{EFT expansions in 3-d}
\label{3d}

We have seen than in 1-d the scale dependence of $T_{\rm 1LPT}^2(k)$
can be fit well with even powers of $k$ and that by expanding $P_{\rm 1LPT}$ 
into an SPT loop series there is a well defined procedure that gives us an 
expansion in both EFT parameters and SPT loop order. As already shown in 
\cite{2012JCAP12011T, 2015arXiv150507098B, 2015arXiv150702255B}, 
in 3-d there is also very little stochasticity between 2LPT and dark matter at 
low $k$, about 10\% in terms of total contribution to the overall nonlinear 
effect for $k<0.2$h/Mpc. For 1LPT the stochasticity is larger, and cannot be 
neglected relative to other nonlinear effects at any $k$. This is similar to the 1-d 
case and has the same origin: 1LPT and 2LPT determine the positions of halo 
formation (where shell crossings occur) well, but LPT displacements do not stop 
there, but instead particles continue to stream and spread the dark matter by a
distance that is approximately the same everywhere, independent of the halo mass. 
In terms of the EFT parameters one therefore expects at the lowest order a similar 
correction as in 1-d case, which is of order of $\alpha_{\rm iLPT}k^2P_{\rm L}(k)$, 
where $\alpha_{\rm iLPT}^{1/2}$ is several Mpc/h and the correction is positive 
relative to 1LPT or 2LPT. In general, we can again define 
\be
P_{\rm dm}=P_{\rm iLPT}\left(1+\alpha_{\rm iLPT}(k)k^2 \right), 
\label{eq:ilpttf}
\ee
where for this paper we have $i=1,2,3$. We will also apply this to CLPTs model 
presented in \cite{2015PhRvD..91b3508V}, where the linear power spectrum is 
truncated at the nonlinear scale. In the following we will use $P_{\rm dm}=
T_{\rm 2LPT}^2P_{\rm 2LPT}$ using $T_{\rm 2LPT}$ from simulations, since 2LPT correlates 
very well with the dark matter and $1-r^2 <1\% $ for $k \sim 0.2$h/Mpc (and much 
less than that for lower $k$ \cite{2012JCAP12011T}). This ensures that sampling 
variance cancels to a large extent, and since $P_{\rm 2LPT}$ can be computed analytically we thus 
have a measurement of $P_{\rm dm}$ without the sampling variance down to 
very low $k$. Our analytic 2LPT power spectrum, $P_{\rm 2LPT}$, is based on the
1-loop LPT calculations presented in \cite{2015PhRvD..91b3508V}.  As it was shown 
there, in 2LPT case, second-order displacement is used to compute the corresponding 
density power spectrum, and since the relation between the density and displacement 
is nonlinear, perturbative analysis relays on the cumulant expansion. Truncation of this 
expansion then implies that 2-loop and higher LPT terms that show up in higher 
cumulants are ignored. 
Nevertheless, this approximation was found to be 1\% accurate against simulations 
for scales $k<0.14$h/Mpc \cite{2015PhRvD..91b3508V}, and above this scale we 
switch from analytic model to the N-body simulation results. 

We performed simulations using wiggle and no wiggle realizations of the same 
initial conditions seeds, using fastPM code \cite{Feng...inprep}.
We construct the no wiggle linear power spectrum from the wiggle power spectrum 
using the method first introduced in in \cite{2015PhRvD..91b3508V} and explained in detail in
appendix \ref{app:nowiggle}, which ensures that these 
two have the same $\sigma_8$, as well as velocity dispersion $\sigma_v$.
For this work, flat $\Lambda$CDM model is assumed with $\Omega_{\rm m}=0.272$,
$\Omega_{\Lambda}=0.728$, $\Omega_{\rm b}/\Omega_{\rm m}=0.167$, $h=0.704$,
$n_s=0.967$, $\sigma_8=0.81$. 
The primordial density field is generated using the matter transfer function by CAMB. 
We ran several simulations of both 1.3Gpc/h and 2.6Gpc/h box size and $2048^3$ particles, 
each with a wiggle and no wiggle initial conditions. 

Next we consider the dependence of $\alpha$ parameter on scale $k$. 
In the case when $\alpha(k)$ shows any scale dependence we call it 
running of $\alpha(k)$ parameter. The motivation is to test these models and 
to determine which one exhibits the least running of $\alpha(k)$ over some range of scales. 
The absence of running improves the predictive value of a given model 
since only a constant parameter needs to be determined while in the case 
of running the full functional form is required (over the same range of scales).
For example, in EFT framework, we expect $\alpha$'s to be constant 
parameters so in this picture observed scale dependence at a given scale 
is the indication of importance of higher order corrections that have not been 
included.

\begin{figure*}[t!]
\centering
\includegraphics[scale=0.366, trim=1.4cm 0 2.5cm 0, clip ]{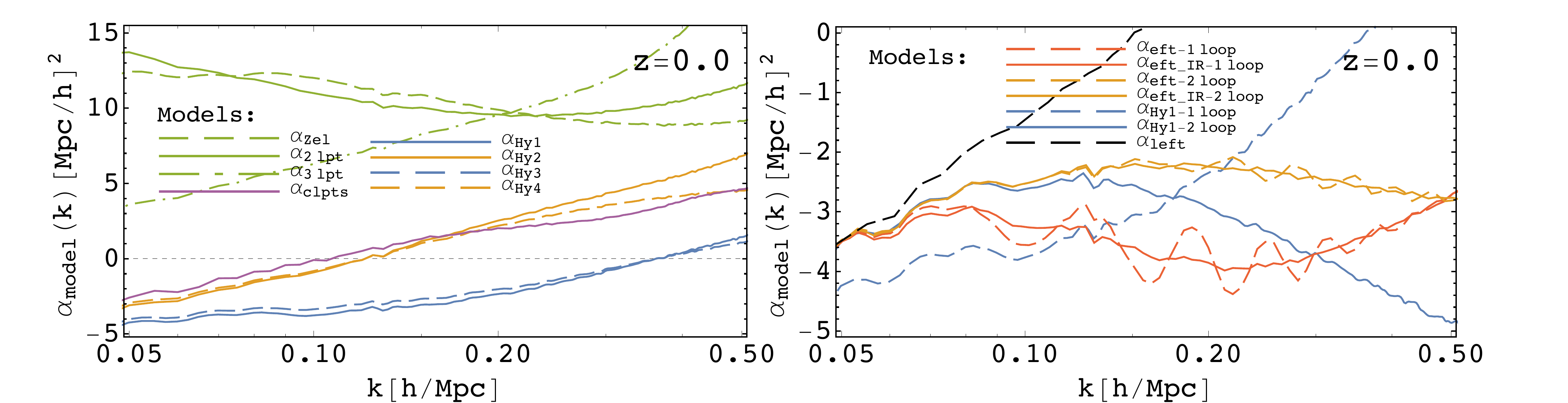}
\caption{
Running of $\alpha(k)$ for severals different models. 
On the \textit{left panel} we show the running of the LPT models (in green) 
related to Eq. \eqref{eq:ilpttf}: 1LPT (solid line), 2LPT (dashed line), 
and 3LPT (dot-dashed line). We also show the CLPTs model in 
Eq. \eqref{eq:clptstf} (purple solid line). On the same panel we show 
the running of the $\alpha$'s related to the hybrid models from Eq. 
\eqref{eq:Hymodels}: Hy1 (blue solid line), Hy2 (orange solid line), 
Hy3 (blue dashed line) and Hy4 (orange dashed line).
On the \textit{right panel} we show one loop (red dashed line) and two loop
(orange dashed line) results for SPT EFT models, and also the IR resummed 
verisins of the same lines (solid red and orange lines).
One loop (blue solid line) and two loop (blue dashed line) Hy1 results also shown, 
as well as one loop results of LEFT \cite{2015JCAP...09..014V}.
}
\label{fig2}
\end{figure*}

In figure \ref{fig2} we present the running of $\alpha_{\rm iLPT}(k)$ as a function of $k$, 
analogous to 1-d analysis (top of figure \ref{fig1}). 
We only plot the results for BAO wiggle 
case, as the wiggle and no wiggle simulations give near identical results for the transfer 
functions. Note that $\alpha_{\rm iLPT}(k)$ show no evidence of BAO features: most of 
the BAO information is thus in $P_{\rm iLPT}$ already. We see 
considerable running of EFT parameters $\alpha_{\rm iLPT}(k)$ over the entire range 
of $k$: we note that even a small amount of running at high $k$ can lead to large effects
in $P_{\rm dm}(k)$.  The accuracy of these measurements depends on the accuracy of 
simulation data. On scales smaller than $k \sim 0.08 h/{\rm Mpc}$ simulation power 
spectrum can be taken as 1\% accurate (see \cite{Feng...inprep} for details).
At low $k$, on the other hand, strong running is exhibited that is most likely caused 
by the inaccuracies in simulations related to the used speedup methods, 
(see \cite{Feng...inprep} for details). 

\subsection{BAO damping derivation}

Recently several resummation procedure have been suggested that include the effects of linear 
displacement field \cite{2015JCAP...02..013S, 2015PhRvD..92d3514B, 2015JCAP...09..014V} on two 
point statistics of wiggles. Such resummations do not improve the reach of the 
perturbative expansions but they do correctly predict BAO damping. 
BAO effects show up in the power spectrum as 2\% residual amplitude oscillations. 
In \cite{2015PhRvD..92d3514B} a simple approximate procedure has been 
presented that captures the bulk of the damping effect on the BAO wiggles. Here we provide an 
alternative derivation. 

\begin{figure*}[tb!]
\centering
\includegraphics[scale=0.56]{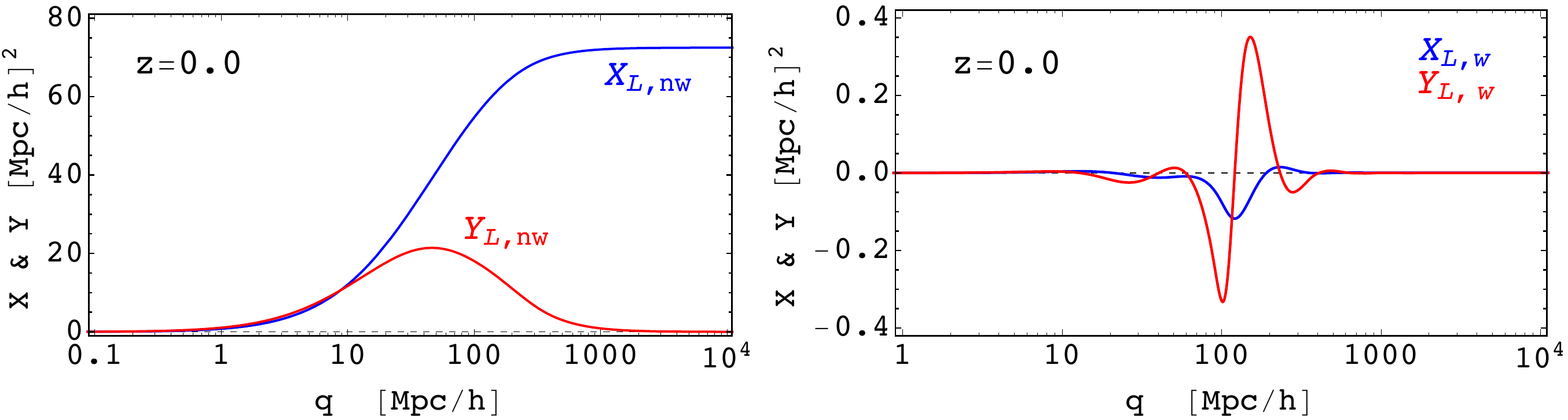}
\caption{
Scale dependence of the linear two point functions of displacement field, 
which contribute to the cumulant expansion, Eq. \eqref{eq:XYex}. 
We have split the contributions into the wiggle (left panel) 
and no-wiggle part (right panel). All results are shown at redshift $z=0.0$.
We see that the wiggle part has a most of the support at scale $\sim$100 Mpc/$h$.
}
\label{figXY}
\end{figure*}

We start by dividing the initial power spectrum into the non-wiggle and wiggle part:
\eeq{
P_{\rm w,L}(k) = P_{\rm nw,L}(k) + \Delta P_{\rm w,L}(k)
}
where $P_{\rm nw}$ it the broad band part and $\Delta P_{\rm w}$ is the residual containing 
only the wiggle part. This split, given above, is not unique which allows us to impose 
further constraints on both $P_{\rm nw}(k)$ and $\Delta P_{\rm w}$. 
We requite that the resulting nonlinear power spectrum obtained form to $P_{\rm L, w}$
and $P_{\rm L, nw}$ give us the same broad band power amplitudes on large and small scales.
This we can achieved by requiring that $\sigma_8$ and $\sigma_v$ be the same 
for the $P_{\rm L, w}$ and $P_{\rm L, nw}$ power spectrum, i.e.
\eeq{
\sigma_8^2 = \int \frac{d^3q}{(2\pi)^3}~W^2(kR)P_{\rm L, w}(k) = 0 ~~ {\rm and} ~~
\sigma_v^2 = \frac{1}{3}\int \frac{d^3q}{(2\pi)^3}~\frac{P_{\rm L, w}(k)}{q^2} = 0,
\label{eq:sigv_sig8}
}
where $W_R(k)=3\left[\sin(kR)/(kR)^3 -\cos(kR)/(kR)^2 \right]$ and $R = 8 {\rm Mpc}/h$.
Detailed construction of the  $P_{\rm nw}$ using the constraints above is given in Appendix \ref{app:nowiggle}.

In Lagrangian formalism the power spectrum can be written as \cite{2015PhRvD..91b3508V,
2015JCAP...09..014V}
\eeq{
 (2\pi)^3\df^D(k)+P(k)=
 \int d^3 q ~e^{-i\vec {q}\cdot\vec {k}} \la e^{-i\vec {k}\cdot\Delta} \ra
 = \int d^3 q ~e^{-i\vec{q}\cdot \vec {k}} \exp \bigg[
 {-\frac{1}{2}k_ik_jA_{ij}(\vec {q})+\ldots} \bigg], 
 \label{eq:PS-nw-nl}
}
where dots $``\ldots"$ represent the three and higher point cumulant contributions, and we have
\eq{
A_{ij}(\vec {q})&=\left\langle\Delta_i\Delta_j\right\rangle_c 
= X(q)\df^K_{ij}+Y(q)\hat{q}_i\hat{q}_j
}
where the two scalar functions are given by
\eq{
  X(q) =& \int_0^\infty \frac{dk}{\pi^2} P_\psi(k) \left[\frac{1}{3} - \frac{j_1(kq)}{kq}\right] , \non \\
  Y(q) =& \int_0^\infty \frac{dk}{\pi^2} P_\psi(k) j_2(kq) ,
  \label{eq:XYex}
}
and $P_\psi(k)$ is the diagonal part of the displacement power spectrum \cite{2015PhRvD..91b3508V}.
In linear approximation we have $P_\psi(k) \to P_{\rm L}(k)$. 
Since $X$ and $Y$ are linearly related to $P_{\rm \psi}$ we can simply separate
\eq{
A_{ij}(q) = A_{ij,{\rm nw}}(q) + \Delta A_{ij,{\rm w}}(q).
}
Keeping only the linear part of $A_{ij}$ exponentiated and expanding the rest we can rewrite
the power spectrum
\eq{
P_{\rm w}(k)&= \int d^3 q ~e^{-i\vec{q}\cdot \vec {k}} e^{-\frac{1}{2}k_ik_jA^{ij}_{\rm L,nw}(\vec {q})} 
\bigg(1 + ``{\rm higher~no-wiggle~terms}" \bigg) \non\\
&~~~~+ \int d^3 q ~e^{-i\vec{q}\cdot \vec {k}} e^{-\frac{1}{2}k_ik_jA^{ij}_{\rm L,nw}(\vec {q})} 
\bigg(-\frac{1}{2}k_ik_j\Delta A^{ij}_{\rm L,w}(\vec {q}) + ``{\rm higher~wiggle~terms}" \bigg)\non\\
&= P_{\rm nw} (k)
+ \int d^3 q ~e^{-i\vec{q}\cdot \vec {k}} e^{-\frac{1}{2}k_ik_jA^{ij}_{\rm L,nw}(\vec {q})} 
\bigg(-\frac{1}{2}k_ik_j\Delta A^{ij}_{\rm L,w}(\vec {q}) + \ldots \bigg)
}
where $``\ldots"$ now represent the terms of one loop order and higher, and $P_{\rm nw}$
is the no-wiggle nonlinear power spectrum given by the same expression as in Eq. \eqref{eq:PS-nw-nl},
with the initial power spectrum $P_{\rm L,nw}$. Expanding the angular part of the $A^{ij}_{\rm L,nw}$
and keeping the monopole part exponentiated we have
\eq{
 \int &d^3 q ~e^{-i\vec{q}\cdot \vec {k}} e^{-\frac{1}{2}k_ik_jA^{ij}_{\rm L,nw}(\vec {q})} 
\bigg[-\frac{1}{2}k_ik_j\Delta A^{ij}_{\rm L,w}(\vec {q}) + \ldots \bigg] = \non\\
&=\int d^3 q ~e^{-i\vec{q}\cdot \vec {k}} e^{-\frac{1}{2}k^2\lb X_{\rm L,nw} + \tfrac{1}{3}Y_{\rm L,nw} \rb} 
\bigg[ -\frac{1}{2}k_ik_j\Delta A^{ij}_{\rm L,w} 
\Big(1 - \tfrac{1}{2}k_ik_j (\hat{q}_i\hat{q}_j -1/3\df^D_{ij})Y_{\rm L,nw} \Big)+ \ldots \bigg] \non\\
&= e^{- k^2 \Sigma^2} \int d^3 q ~e^{-i\vec{q}\cdot \vec {k}} \bigg[-\frac{1}{2}k_ik_j\Delta A^{ij}_{\rm L,w} 
+ \frac{1}{6}k^2 k_ik_j \Delta A^{ij}_{\rm L,w} \mathcal P_2(\mu)Y_{\rm L,nw} + \ldots \bigg] \non\\
&= e^{- k^2 \Sigma^2} \Big( \Delta P_{\rm L,w} + ``{\rm higher~order~wiggle~terms}" \ldots \Big),
\label{eq:IRder}
}
where $\mathcal P_2(\mu)$ is the second Legendre polynomial. We eventually neglect the term 
proportional to the $\mathcal P_2(\mu)$ angular dependence which, at the leading order, leaves 
only first term in the squared brackets.
We have also introduced averaged quantity $\Sigma^2 = \tfrac{1}{2}\big\llangle X_{\rm L,nw}(q) 
+ \tfrac{1}{3}Y_{\rm L,nw}(q) \big\rrangle_{\rm w} \simeq 28.0 ({\rm Mpc}/h)^2 $ where $\llangle \cdot \rrangle$ 
represents the averaging over $q$ range where support of $\Delta A^{ij}_{\rm L,w}$ is prominent, 
i.e. $q\simeq 110 {\rm Mpc}/h$. In Eq. \eqref{eq:IRder} above we have used the fact that in the region 
where $A^{ij}_{\rm L,w}$ has a non-negligible support (around the scale of $\sim 110 {\rm Mpc}/h$), $A^{ij}_{\rm L,nw}$ 
varies slowly. This can be seen in figure \ref{figXY} where $X$ and $Y$ components of both wiggle and 
non-wiggle part are shown. This approximation also known as Laplace's approximate integration method.
We note that fully nonlinear $X$ and $Y$ are nearly identical to the linear 
around $q\simeq 110 {\rm Mpc}/h$ \cite{2014JCAP...06..008T}.
It has been argued \cite{2015JCAP...02..013S} that this IR resummation does not affect the 
broadband SPT or EFT terms, so we will leave the broadband part unchanged. 
Thus our full IR-SPT model at one loop gives
\begin{align}
P_{\rm dm}(k)&=P_{\rm nw,L}(k)+P_{\rm nw, SPT,1-loop}(k)
+\alpha_{\rm SPT,1-loop,IR}(k)k^2P_{\rm nw,L}(k) \\
&\hspace{1.8cm}+ e^{-k^2\Sigma^2} \Big(\Delta P_{\rm w,SPT,1-loop } (k)+
\left(1+(\alpha_{\rm SPT,1-loop,IR}+\Sigma^2)k^2\right)\Delta P_{\rm w,L}(k)\Big).\nonumber
\label{eq:eft1loop}
\end{align}
where we have introduced one-loop wiggle only power spectra
\begin{align}
\Delta P_{\rm w,SPT,1-loop } (k)& = P_{\rm w,SPT,1-loop}(k)-P_{\rm nw,SPT,1-loop}(k).
\end{align}
Similar procedure can also be straightforwardly applied to the hybrid and 2-loop 
SPT results as discussed further below (see also \cite{2015arXiv150702255B}). 
While we have focused on BAO wiggle here, the derivation applies to any wiggle localized
in $q$. In general, the lower value of $q$ the lower the value of damping distance $\Sigma$. 

\subsection{Power spectrum results}

We observe that the running for 1LPT and 2LPT is larger than in the 1-d case. The main 
difference between 3-d and 1-d analysis is that for the latter the full displacement 
solution is given by the first order 1LPT (Zeldovich) displacement ${\bf\Psi}_1$, while 
in 3-d one has an infinite series of displacements, 
${\bf\Psi}=D_+{{\bf\Psi}_1}+D_+^2{\bf\Psi}_2+D_+^3{\bf\Psi}_3+...$. At 1-loop SPT order 
in $P(k)$, we have contributions up to 3LPT (${\bf\Psi}_3$). But this term is not included
in $P_{\rm 2LPT}$, which therefore does not contain the full 1-loop SPT. Similarly, 1LPT 
does not include both ${\bf\Psi}_2$ and ${\bf\Psi}_3$. If, as expected, the correct dark 
matter solution contains the full SPT 1-loop terms, then the running of 
$\alpha_{\rm 1LPT}(k)$ and $\alpha_{\rm 2LPT}(k)$ reflects the $k$ dependence of 
the difference between 1LPT and 2LPT at 1-loop level versus the full 1-loop SPT. 
In contrast, 3LPT contains all 1-loop SPT terms and we expect the running of 
$\alpha_{\rm 3LPT}(k)$ to have less $k$ dependence at low $k$. However, 3LPT 
contains additional terms beyond 1-loop SPT that are quite large and largely spurious 
\cite{2015PhRvD..91b3508V}, so we do not expect 3LPT EFT 
parameter to agree with the corresponding SPT 1-loop EFT parameter defined below. 
For example, at low $k$ 3LPT has an additional zero lag contribution 
$k^2P_{\rm L}(k)\sigma^2_{13}$, which is a 2-loop SPT term, and is quite large. 
At higher $k$ there are additional 2-loop SPT terms in 3LPT, which cause 
$\alpha_{\rm 3LPT}(k)$ to be running with $k$. 

To gain more insight into the running of these parameters 
we can perform the expansion of 1-3LPT at low $k$, which has been shown to be 
accurate for $k<0.12$h/Mpc \cite{2015PhRvD..91b3508V}. We have 
\begin{align}
P_{\rm SPT,1-loop}(k)& =\big(1-k^2\sigma_{\rm L}^2\big)P_{\rm L}(k) +\frac{1}{2}Q_3(k)+\frac{9}{98}Q_1(k)
                                 +\frac{10}{21}R_1(k) + \frac{3}{7}\Big(Q_2(k)+2R_2(k)\Big), \nonumber \\
P_{\rm 1LPT,1-loop}(k)&=\big(1-k^2\sigma_{\rm L}^2\big)P_{\rm L} (k)+\frac{1}{2}Q_3(k),\nonumber\\
P_{\rm 2LPT,1-loop}(k)&=\big(1-k^2\sigma_{\rm L}^2\big)P_{\rm L} (k)+\frac{1}{2}Q_3(k)
                                   +\frac{3}{7}\Big(Q_2(k)+2R_2(k)\Big)+\frac{9}{98}Q_1(k), \nonumber \\
P_{\rm 3LPT,1-loop}(k)&=P_{\rm SPT,1-loop}(k)-k^2P_{\rm L}(k)\sigma^2_{\rm 1loop},
\label{lowklpt}
\end{align}
where we have introduced several LPT terms defined in e.g. \cite{2008PhRvD..77f3530M}. 
However, the expression for $P_{\rm 3LPT,1-loop}$ is not necessarily valid at low $k$. 
The reason is that there is a large zero lag correlation of 22 and especially of 13 
displacements. At the next order in 3LPT we have
$P_{\rm 3LPT}=P_{\rm SPT,1-loop}(k)-k^2P_{\rm L}(k)\sigma^2_{\rm 1loop}$, 
where $\sigma^2_{\rm 1loop}$ is the sum of zero lag correlations of 22 and 13 LPT 
displacements. At $z=0$ the linear theory value of $\sigma_{\rm L}^2 \sim 36~(\text{Mpc}/h)^2$,
and $\sigma_{\rm 1loop}^2=7~(\text{Mpc}/h)^2$, so the 2-loop SPT term in 3LPT 
is quite large even at very low $k$. It is almost entirely spurious \cite{2015PhRvD..91b3508V} 
(see also \cite{2015arXiv150702255B} for recent results). In \cite{2015PhRvD..91b3508V} we 
introduced CLPTs model that attempts to cure this problem by truncating high $k$ contributions.
We can define a model related to the CLPTs in a similar way as in Eq. \eqref{eq:ilpttf}
\be
P_{\rm dm}=P_{\rm CLPTs}\left(1+\alpha_{\rm CLPTs}(k)k^2 \right), 
\label{eq:clptstf}
\ee
Results for this model are also shown in figure \ref{fig2}. We observe that EFT term is close to 0
around $k \sim 0.1$h/Mpc, but it still has a strong $k$ dependence. 

We see that the 2LPT solution does not contain all the terms at the 1 loop SPT level. 
Why is this not important for its correlation with full dark matter, i.e. why can we drop 
1-loop SPT term in 2LPT and still have near perfect correlation with the final dark matter? 
This is because the missing term is dominated by $R_1(k) \sim \alpha_{R_1}(k)k^2P_{\rm L}(k)$. 
Since we are allowing a fully general $\alpha_{\rm 2LPT}(k)$, it can include this term running 
with $k$. Since $\alpha_{R_1}(k)$ is strongly changing with $k$ even for $k<0.1$h/Mpc, 
it leads to a strong running of $\alpha_{\rm 2LPT}(k)$. Figure \ref{fig2} shows there 
is also considerable running of 3LPT, even though it contains all 1-loop SPT  terms. 
This suggests that 2-loop and higher order contributions cannot be neglected 
except at very low $k$, as also shown in \cite{2015PhRvD..91b3508V}. 
These higher order terms may however be spurious. 

For the SPT perturbative expansion we may expect EFT parameter to be 
running less at low $k$, just as in 1-d case, at least in the range where 1 loop SPT is 
expected to be valid. 
We can define 1 loop SPT EFT parameter $\alpha_{\rm SPT,1-loop}$ as \cite{2012JHEP...09..082C}
\be
P_{\rm dm}(k)=D_+^2P_L(k)\left[1+\alpha_{\rm SPT,1-loop}(k)k^2\right]+D_+^4P_{\rm SPT,1-loop},
\ee
and derive the running of $\alpha_{\rm SPT,1-loop}(k)$ from it. 
The leading EFT terms scale with $P_{\rm L}(k)$, and hence contain BAO wiggles. 
It has been shown that this SPT EFT version is a poor model in the context of 
BAO wiggles \cite{2012JHEP...09..082C}. 
At higher $k$ the higher order effects of long wavelength modes leads to 
suppression of these BAO wiggles \cite{2015JCAP...02..013S}. 
This resummation procedure is automatic in LPT schemes, which are therfore superior 
for BAO damping. 

Figure \ref{fig2} suggests that $\alpha_{\rm SPT,1-loop,IR}(k)$ is nearly constant 
for $k<0.1{\rm h/Mpc}$, with a value around -2$({\rm Mpc/h})^2$. This value increases 
to about -4$({\rm Mpc/h})^2$ for $k>0.1{\rm h/Mpc}$. This change 
suggests that the value of $\alpha_{\rm SPT,1-loop,IR}$ for $k>0.1{\rm h/Mpc}$ is no 
longer the low $k$ EFT parameter at the lowest order. 
We will test this model further using BAO oscillations in next section. 

In 1-d we have identified LPT approach as clearly superior to SPT approach because 
of the automatic resummation of all SPT terms. Doing SPT loops has no advantages 
over 1LPT, since only in the infinite loop limit it converges to 1LPT, and the convergence 
is slow. In 3-d this is no longer the case, because higher order LPT displacements become
less and less reliable, and their resummation is not necessarily a good thing. However, 
doing resummation on 1LPT only is still likely to be useful, since it is dominated by 
modes in the linear regime, hence reliably computed by 1LPT resummation. 
In 3-d, 1-loop SPT identifies properly all 1-loop terms and is not contaminated by 2 and 
higher loop orders, and as a consequence we expect it to give a constant 
EFT parameter over the range of its validity and over the range of constant 
$\alpha$ EFT parameter validity. Moreover, we expect the EFT parameter relative 
to 1-loop SPT to be small \cite{2012JHEP...09..082C}, and indeed the values in 
figure \ref{fig2} suggest it is a lot smaller than the corresponding LPT versions.  

We thus want a scheme where we resum only 1LPT (or 1LPT and 2LPT), and add 
the remaining SPT 1-loop terms, and finally add or multiply EFT terms to it. 
We can define several versions of this proposal, 
\begin{align}
\text{Hy1:}~~  P_{\rm dm}(k)&=P_{{\rm 1LPT}}(k)\left(1+\alpha_{\rm 1LPT,1-loop}(k)k^2 \right)
+\Big(P_{\rm SPT,1-loop}(k)-P_{\rm 1LPT,1-loop}(k)\Big)_{IR} \nonumber\\
\text{Hy2:}~~ P_{\rm dm}(k)&=P_{{\rm 2LPT}}(k)\left(1+\alpha_{\rm 2LPT,1-loop}(k)k^2 \right)
+\Big(P_{\rm SPT,1-loop}(k)-P_{\rm 2LPT,1-loop}(k)\Big)_{IR} 
\nonumber \\
\text{Hy3:}~~ P_{\rm dm}(k)&=P_{{\rm 1LPT}}(k)+\Big(P_{\rm SPT,1-loop}(k)-P_{\rm 1LPT,1-loop}(k)
+P_L\alpha_{\rm SPT,1LPT,1-loop}(k)k^2\Big)_{IR} . 
\label{eq:Hymodels}
\end{align}
All these models can readily be derived from a full resummed form of the power spectrum 
given in Eq. \eqref{eq:PS-nw-nl}. The difference is only on which terms we keep exponentiated 
and which we expand. In the case of Hy1 model 1LPT displacement (Zel'dovich part) 
remains resummed while rest of the one-loop contribution is expanded. Similarly for Hy2 
model we leave 2LPT displacement contributions resummed while we expand the one-loop 
residual. The difference of Hy1 and Hy3 models is in the way contributions proportional to 
$\alpha$ is resummed; in Hy1 case, it is proportional to Zel'dovich contribution and in Hy3 
case to the linear power spectrum.
These results are also shown in figure \ref{fig2} and show considerable running of EFT parameters.
Models Hy1 and Hy3 are almost equal, and we will show below they give very similar results. 
The same is also true for recently developed LEFT \cite{2015JCAP...09..014V}, 
which as well shows substantial running of $\alpha(k)$ at all $k$.

In 1-d we have seen that SPT expansion have a smaller radius of convergence 
than the EFT expansion, so to improve the model we could repeat the EFT 
procedure on 2-loop SPT even with just 1 EFT parameter. We can assume that 
1-loop SPT EFT parameter remains unchanged. Unlike the 1-d case, we must 
also absorb with the EFT parameter the very large zero lag value of 
$\sigma_{\rm 1loop}^2$, which contributes at 2 loop order a term 
$-k^2\sigma^2_{\rm SPT,2-loop}P_L(k)$, which dominates at low $k$, 
but which is almost entirely spurious \cite{2015PhRvD..91b3508V, 
2015arXiv150702255B}. However, we 
would have to also absorb its 2-loop counter terms. It is clear that doing 
2-loop SPT is considerably more complicated than in 1-d. 
If we assume low $k$ 2-loop SPT is completely spurious at 
low $k$ this is equivalent to the requirement that 2-loop SPT vanishes at 
low $k$, just as in 1-d case, 
\begin{align}
P_{\rm dm}(k)=\big(1+\alpha_{\rm SPT,1-loop}(k)k^2 \big) P_{\rm L}(k)
&+\big( 1+\alpha_{\rm SPT,1-loop}(k)k^2 \big) P_{\rm SPT,1-loop}(k) \nonumber\\
&+\big( P_{\rm SPT,2-loop}(k)-k^2\sigma_{\rm SPT,2loop}^2 P_{\rm L}(k)\big).
\label{eq:eft2loop}
\end{align}
Result for the running of $\alpha$ for this model is shown in figure \ref{fig2}.
More generally, if we split $\sigma_{\rm SPT,2loop}^2$ into a spurious part and a real part then one finds
one needs a different EFT term multiplying $P_{\rm L}$ and $P_{\rm SPT,1-loop}$,
\be
P_{\rm dm}(k)=\big(1+\alpha_{\rm SPT,2-loop}(k)k^2\big)P_{\rm L}(k)
+\big(1+\alpha_{\rm SPT,1-loop}(k)k^2\big)P_{\rm SPT,1-loop}(k)
+P_{\rm SPT,2-loop}(k).
\ee
In 1-d we have seen that at 2-loop level one can assume 
$\alpha_{\rm SPT,2-loop}(k)=\alpha_{\rm SPT,1-loop}(k)$ and that one does not need 
to add its $k^2$ dependence, but in principle one could add that as another parameter. 
This has been explored in \cite{2014JCAP...07..057C}, but we do not pursue it further here. 

\begin{figure*}[tb!]
\centering
\includegraphics[scale=0.5]{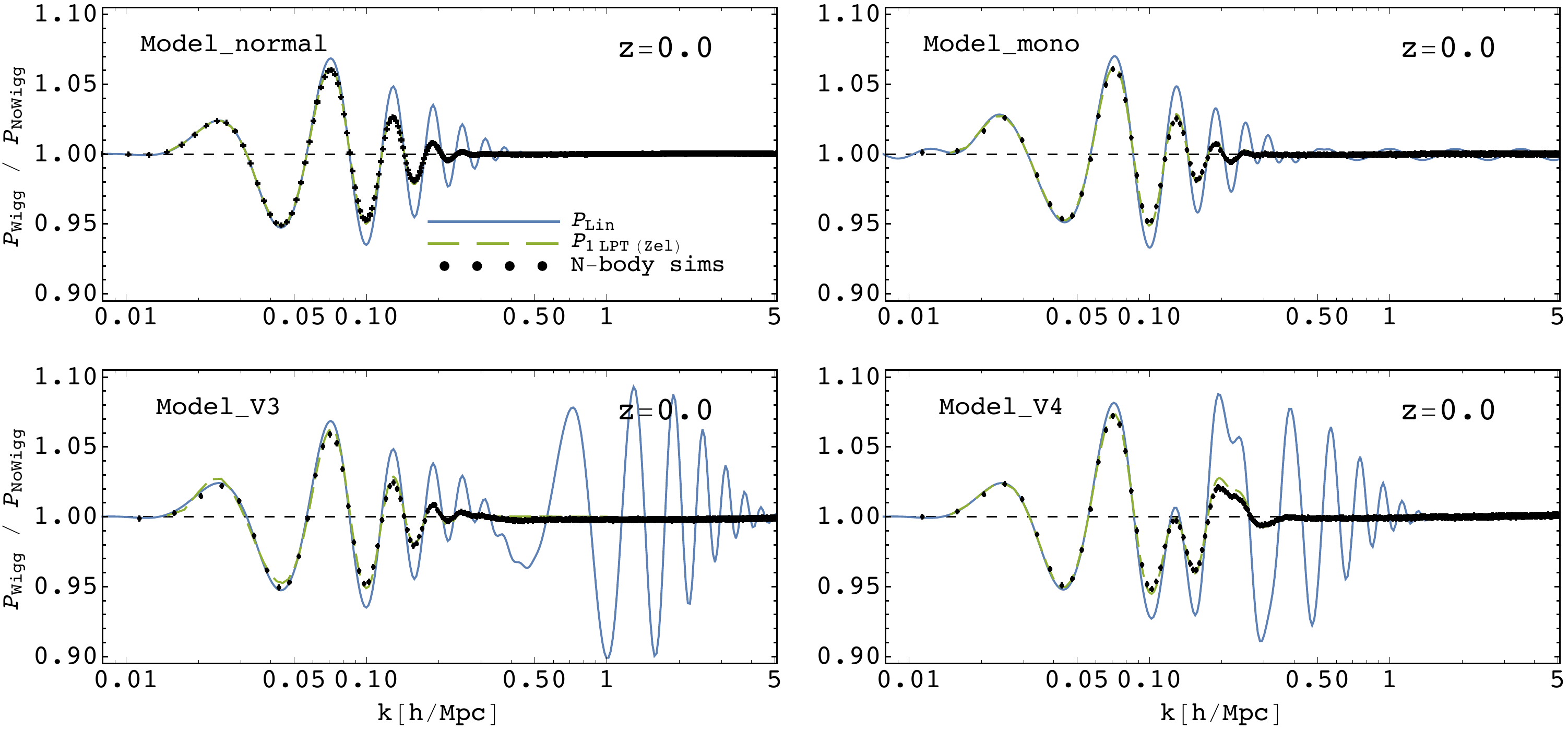}
\caption{
BAO wiggles, i.e. ratio of the wiggle and non-wiggle power spectrum, is shown for 
four different models: $\Lambda$cdm (top left panel), Monodromy model 
(see \cite{2010JCAP...06..009F, 2014arXiv1406.0548M}) (top right panel) and two other models labeled
$V_3$ and $V_4$ with additional wiggles relative to $\Lambda$cdm (bottom panels).
Linear theory results (blue lines) evolve due to nonlinearities and yield results 
given by N-body simulations (black points). Wiggle damping for all these models 
is well described by the 1LPT (Zel'dovich) model (green dashed line). 
Note that all the initial wiggles are highly dampened at lower scales, $k \lesssim 0.5 h/$Mpc.
All results are shown at redshift $z=0.0$.
}
\label{fig3}
\end{figure*}

\subsection{Higher order correlations} 

So far our approach has been to quantify the ignorance of PT by the transfer 
function, which can be translated into an effective EFT parameter that is fitted 
to simulations. Over a narrow range of $k$ this is guaranteed to give correct 
answer, and hence the method is simply parameterizing the ignorance with 
essentially no physics. However, the concept of transfer functions in the absence 
of stochasticity can be more useful than that: it allows one to derive the higher 
order correlations in the regime where $r_{\rm 2LPT} \sim 1$. For example, 2LPT 
already contains all of the terms that determine density perturbation at 2nd order, 
\be
\delta_{\rm 2LPT}(\vk)=\delta_{\rm L}(\vk)
+\int d^3\vk_1 d^3\vk_2 F_2(\vk_1,\vk_2)
\delta^D({{\bf k}-{\bf k}_1-{\bf k}_2})\delta_L(\vk_1)\delta_L(\vk_2),
\ee
where $F_2(\vk_1,\vk_2)$ is the SPT 2nd order kernel 
(see e.g. \cite{2002PhR...367....1B}), 
\be
  F_2(\vk_1,\vk_2) = \frac{17}{21} + \frac{1}{2}\left(
\frac{k_1}{k_2}+\frac{k_2}{k_1}
\right) \hat\vk_1\cdot\hat\vk_2
+
\frac{2}{21}\left(3(\hat\vk_1\cdot\hat\vk_2)^2 -1\right),
\label{F21}
\ee
where $\hat \vk_i=\vk_i/k_i$.
Bispectrum is defined as 
\be
(2\pi)^3\delta^D({\bf k}_1+{\bf k}_2+{\bf k}_3) B({\bf k}_1,{\bf k}_2,{\bf k}_3)=
\langle \delta({\bf k}_1)\delta({\bf k}_2)\delta({\bf k}_3) \rangle,
\ee
and hence, at the tree level, one can write
\be
B({\bf k}_1,{\bf k}_2,{\bf k}_3)=
T(k_1)T(k_2)T(k_3)\Big( F_2(\vk_1,\vk_2)P_{\rm L}(k_1)P_{\rm L}(k_2)
+ {\rm 2\, cycl.\, perm.} \Big).
\ee
At the lowest order we can write $T(k)=(1+\alpha_{\rm 2LPT}(k)k^2)^{1/2} 
\sim 1+\alpha_{\rm 2LPT}(k)k^2/2$ and thus the above expression becomes
\ba
B({\bf k}_1,{\bf k}_2,{\bf k}_3)=
\Big(1+\alpha_{\rm 2LPT}(k_1)k_1^2/2&+\alpha_{\rm 2LPT}(k_2)k_2^2/2
+\alpha_{\rm 2LPT}(k_3)k_3^2/2\Big)  \non\\
&\times\Big[F_2(\vk_1,\vk_2)P_{\rm L}(k_1)P_{\rm L}(k_2)+{\rm 2\, cycl.\, perm.}\Big].
\ea
To this one needs to add 1-loop bispectrum from 2LPT terms, which are 
at the same order as the EFT corrections of tree level bispectrum. 
This requires expanding $\delta_{\rm 2LPT}$ to 4th order in $\delta_{\rm L}$. 
The 2LPT 1-loop bispectrum contains some, but not all of the terms of 
the full SPT bispectrum at the 1-loop order. Some of the missing terms may 
cancel out the scale dependence of $\alpha_{\rm 2LPT}(k)$ in the EFT corrected 
tree level bispectrum above, but at the operational level this is not relevant: 
the expression above should include all of the terms at 1-loop level, including 
the EFT terms and should be valid as long as $r_{\rm 2LPT}=1$. As mentioned 
above stochasticity in 2LPT has about 10\% contribution to the nonlinear 
terms for $k<0.2$h/Mpc, and rapidly grows above that, so the expressions 
above should only be valid at this level. Previous analyses \cite{2015JCAP...05..007B,
2014arXiv1406.4143A} have argued 
that there are 4 different EFT parameters (although only 1 matters for improving 
the fits), while in the expression above there is only the transfer function, 
which is the same that also enters into the power spectrum calculations. 
While this transfer function, when expressed in EFT terms $\alpha_{\rm 2LPT}(k)$ 
runs with $k$ and cannot be approximated as a constant, one can derive this 
running from the power spectrum. As long as we work up to the same $k$, 
and stochasticity can be neglected, there are no additional EFT parameters 
for bispectrum relative to the power spectrum. At the leading order any EFT 
corrections can only be a multiplicative factor times $F_2$ kernel, analogous 
to the power spectrum situation, where at the leading order one can only have 
EFT corrections multiplying the linear power spectrum $P_L(k)$. This may help 
explain why the additional parameters in EFT fits of bispectrum were found not 
to be needed \cite{2015JCAP...05..007B,2014arXiv1406.4143A}. Same concepts 
can be applied to higher order correlators (trispectrum etc.) as well. We do not pursue 
this approach further here, but it would be interesting to see how it compares against 
the standard EFT bispectrum calculations \cite{2015JCAP...05..007B,2014arXiv1406.4143A}.

\section{BAO residuals in EFT schemes}
\label{bao}

\begin{figure*}[tb]
\centering
\includegraphics[scale=0.39, trim=0 0 1.4cm 0, clip ]{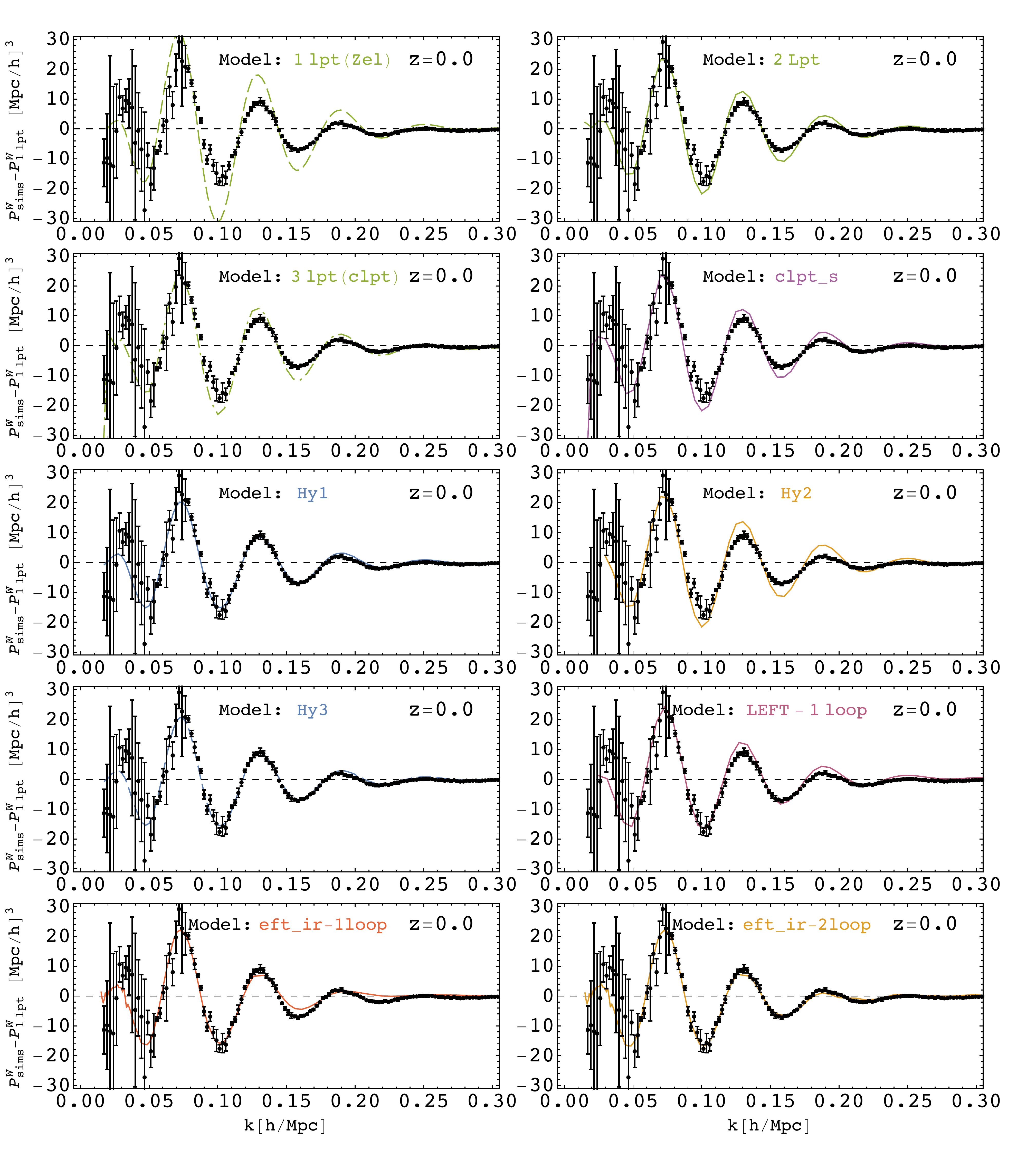}
\caption{
Residual wiggles, relative to the 1LPT (Zel'dovich). In top two lines iLPT models, 
as well as CLPTs model (see \cite{2015PhRvD..91b3508V}) are shown using 
definitions in Eq. \eqref{eq:ilpttf}.  In panels in lines three and four we show 
residuals of hybrid models defined in Eq.  \eqref{eq:Hymodels}, as well as 
LEFT theory developed recently in \cite{2015JCAP...09..014V}. 
In a bottom line we show SPT-EFT one loop and two loop 
models (IR resummation included), given by Eq. \eqref{eq:eft1loop}
and \eqref{eq:eft2loop} respectively. All results are shown at redshift $z=0.0$.
}
\label{fig4}
\end{figure*}

Armed with these various EFT expansions we can test them on BAO wiggles. 
Let us first focus on the BAO residuals in standard $\Lambda$cdm model. 
The main idea is that EFT terms scale with $P_{\rm L}(k)$, and hence contain 
BAO wiggles, and the better the scheme the better it should be able to explain 
the BAO wiggles. We compute the power spectrum for both wiggle $P_{\rm w}(k)$ 
and no-wiggle $P_{\rm nw}(k)$ simulations and take the ratio of the two. 
This is shown in figure \ref{fig3} top left panel, together with linear theory and 
Zeldovich approximation (1LPT). It is clear that 1LPT does a good job in 
describing the BAO damping. In the same figure we explore also the wiggles 
of some other (non-BAO/$\Lambda$cdm) models, which are discussed further below.  

Next, we look at the small wiggle residuals. 
In order to do this subtract out the 1LPT part and then compare the 
models and N-body simulations. 
This is shown in figure \ref{fig4}. We see that there is a residual BAO, 
even if it is small, about a factor of 10 smaller than the original BAO. 
It is these residuals that we wish to test against the EFT expansions. 
Our model for BAO residuals is simply given by $P_{\rm dm,w}(k)$ and 
$P_{\rm dm,nw}(k)$ versions of the EFT models presented in previous 
section. EFT of 1LPT \footnote{We note that our 1LPT  EFT model is 
equivalent to the ZEFT model introduced in $\left[25\right]$.}
is a decent model and predicts some of the residuals, 
but 2LPT and 3LPT (all defined by Eq. \eqref{eq:ilpttf}) are a lot better. 
For 1LPT the stochastic term is not negligible \cite{2012JCAP12011T}, 
so we expect a worse agreement if we apply $\alpha_{\rm 1LPT}(k)$ to 
model BAO residuals. In the same family we can also add the CLPTs model 
(see \cite{2015PhRvD..91b3508V} ) for which EFT extension is defined in a same 
way as it was for iLPT models, Eq. \eqref{eq:ilpttf}. Performance of CLPTs on the wiggle 
residuals is of the same level of accuracy as 3LPT.
Next, we look at the hybrid models defined in Eq \eqref{eq:Hymodels}. 
The difference of these models are in resummation of up to 1LPT terms 
(for Hy1 and Hy3 models) or up to 2LPT terms (for Hy2). 
These models can be considered to be at 1-loop level (up to the resumation). 
We see that 1LPT resummation models (Hy1 and Hy3) work better on the 
wiggle residuals then 2LPT (Hy2). 
Indeed residuals of Hy1 and Hy3 are amongst the best models 
considered and their performance is matched only by two-loop SPT model. 
LEFT \cite{2015JCAP...09..014V} 
performs at the similar level of accuracy as IR resummed one-loop SPT.
For one-loop SPT we also find a decent fit for IR versions, and not very good 
fit for non-IR versions (as can be already seen from figure \ref{fig2}). Going to 
two-loops IR resummed SPT improves the wiggle residuals further 
and gives a very good agreement.  
Thus our results suggest that these residuals can be modelled as a smooth function 
multiplying PT power spectrum. These results suggests that the broadband 
analysis extracting transfer functions relative to LPT or SPT is also able to 
reproduce the BAO wiggles, so the picture is consistent, and in all cases introducing 
EFT parameters improves the agreement on the wiggle part of the power spectrum. 

\subsection{Wiggles in primordial power spectrum}

Here we focus on the wiggles and wiggle-like features in the power spectrum 
beyond the $\Lambda$cdm. These could be imprinted by the physical processes 
during inflation, for example. It is interesting to see how the models discussed 
above perform in predicting such features given that $\alpha(k)$ has been 
determined from the broad band spectra. For this purpose we construct three 
new wiggle models; first is the monodromy-like model (see \cite{2010JCAP...06..009F} 
and \cite{2014arXiv1406.0548M} for the parameteization we have adopted), 
which has oscillations in $\ln k$, 
and the other two models (labeled $V_3$ and $V_4$) 
are using the same BAO wiggle power spectrum with the boosted amplitude and 
shifted scale dependence. These additional wiggle power spectra are then added 
to the $\Lambda$cdm power spectrum. 
In figure \ref{fig3} we show the ratio of the total ($\Lambda$cdm plus the additional wiggles) wiggle 
to non-wiggle power spectrum. Linear theory shows the features of the initial power spectrum 
which are then subject to the nonlinear evolution. We can see that all the initial wiggles are 
highly damped at high $k$, $k > 0.5 h/$Mpc, in fact, wiggles of $V_3$ model 
are completely washed our at redshift $z=0.0$.

\begin{figure*}[tb!]
\centering
\includegraphics[scale=0.506]{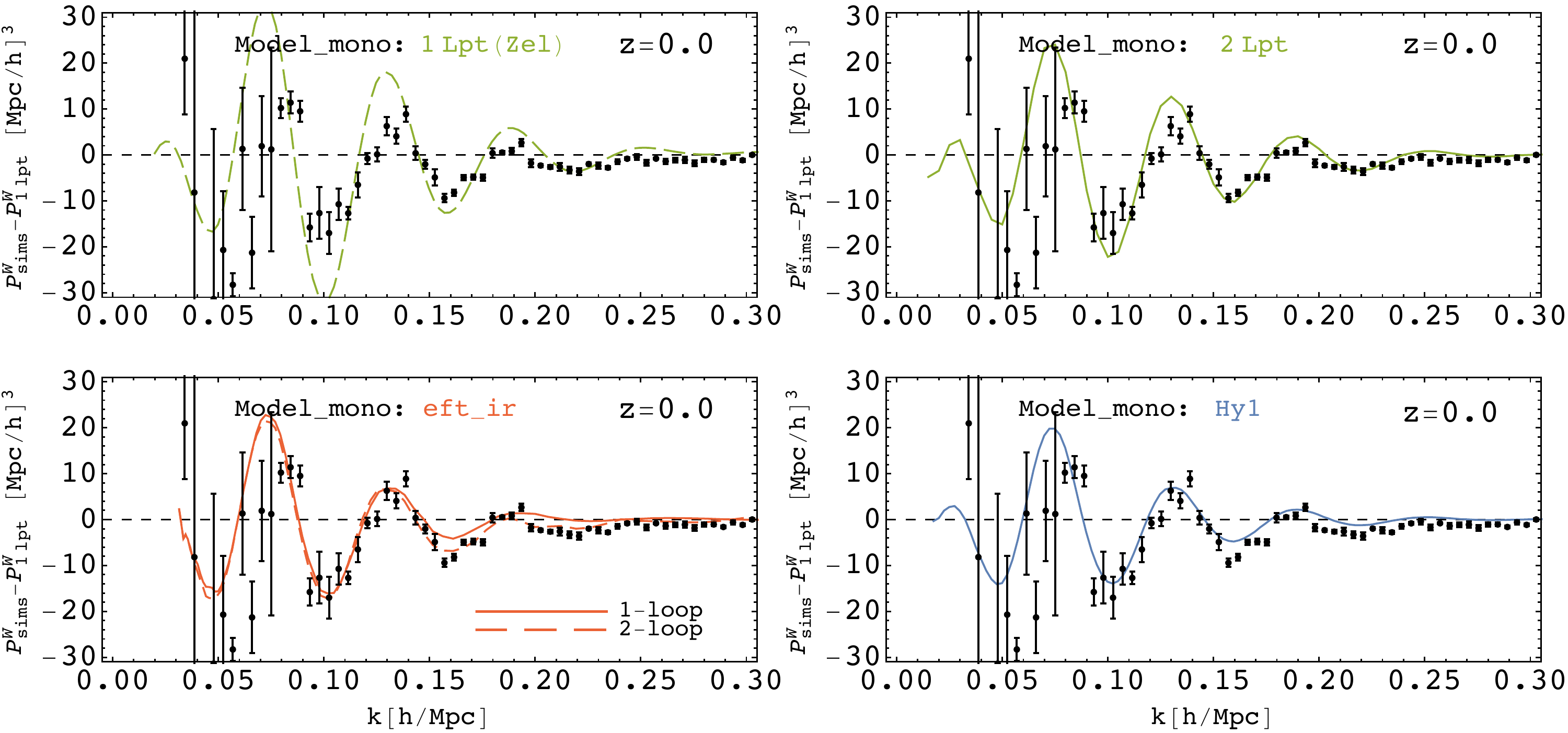}
\caption{
Residual wiggles for the monodromy model shown in \ref{fig3}, relative to the 1LPT (Zel'dovich).
We compare the performance of transfer function extended of 1LPT (green dashed line)
as well as 2LPT  (green solid line) models (given by Eq. \eqref{eq:ilpttf}). In addition 
we show the performance of one loop (red solid line) and two loop (red dashed line)
SPT models as well as one loop Hy1 model (see Eg. \eqref{eq:Hymodels}).
Note that the same values of $\alpha(k)$ are used as in $\Lambda$ model.
Results are shown for redshift $z=0.0$.
}
\label{fig5}
\end{figure*}

\begin{figure*}[tb!]
\centering
\includegraphics[scale=0.506]{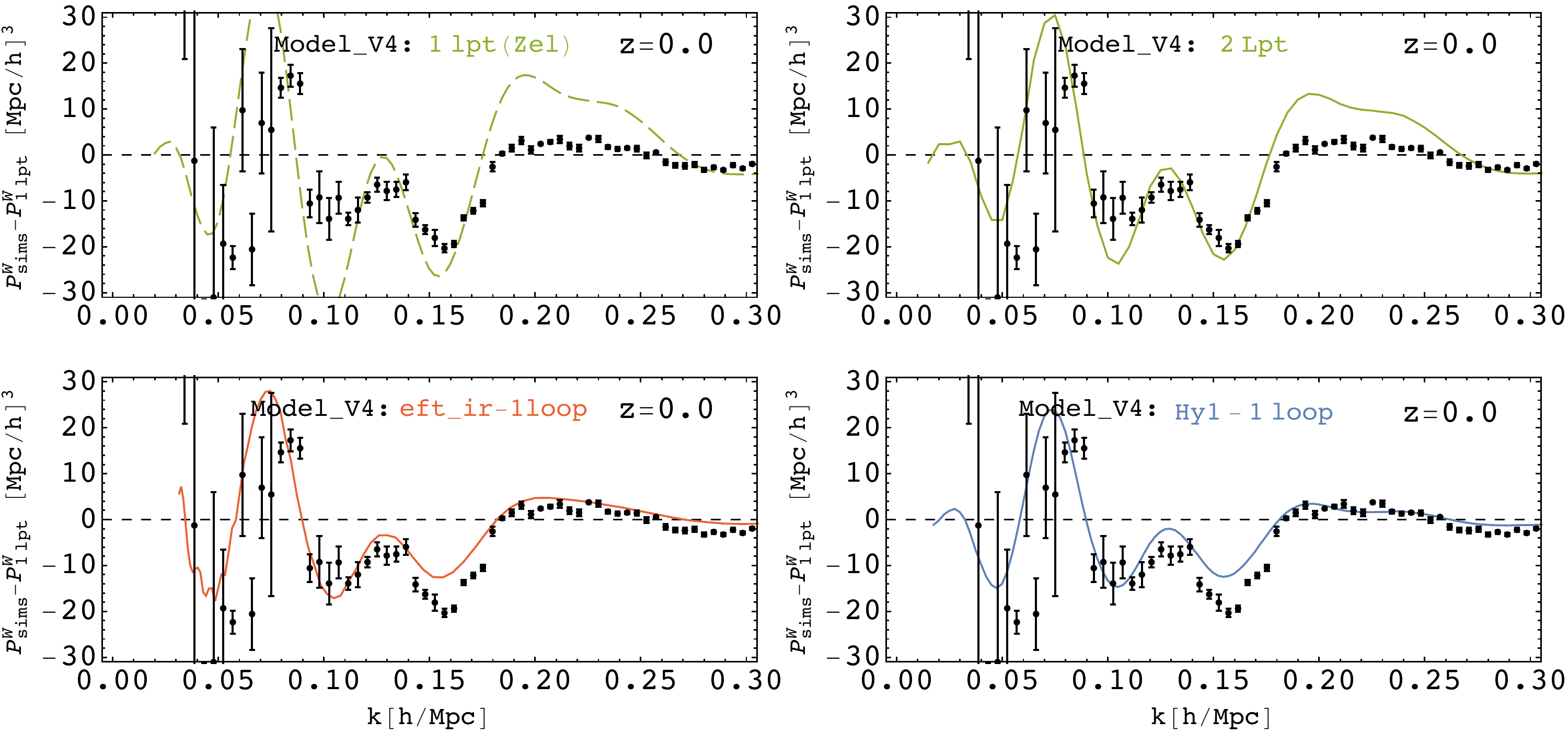}
\caption{
Same as figure \ref{fig5}, for the model $V_4$ shown in figure \ref{fig3}.
}
\label{fig6}
\end{figure*}

We then again look at the small wiggle residuals for these models, where we  
subtract out the 1LPT part and then compare the models and N-body simulations. 
This is shown in figure \ref{fig5} for monodromy-like model and in figure \ref{fig6} for $V_3$
model (as $V_4$ residuals are equivalent to the $\Lambda$cdm at $z=0.0$ we do not show them).
In addition to the 1LPT residuals we show the 2LPT, EFT-SPT and Hy1 model residuals. 
We see that as in the earlier case EFT-SPT and Hy1 perform very well and 
reproduce the wiggle shape to high accuracy. 

\section{Interpreting the dark matter power spectrum}
\label{pk}

In this paper we have addressed the EFT modeling of power spectrum, 
which introduces EFT parameters that parametrize the ignorance of PT and 
at low $k$ scale as $k^2$ multiplied by a low order PT. 
These have to be 
supplemented by stochastic terms at higher $k$. 
In \cite{2014MNRAS.445.3382M, 2015PhRvD..91l3516S} a halo inspired model 
for the dark matter clustering was advocated, where one takes 1LPT (Zel'dovich) 
for the 2-halo term, and adds to it an effective 1-halo term that is very localised 
(to a few Mpc/h). It has been shown that one can build a model with high accuracy 
using this kind of ansatz. The 1-halo term is not necessarily defined as the mass 
within the virial radius: there is nonlinear clustering outside the virial radius that 
adds to this effective 1-halo term. In this section we look at the relation between the 
two approaches. 

Both 1-d and in 3-d LPT simulations suggest that 1LPT (and higher order LPT) 
temporarily create dark matter halos in the regions of orbit crossings, but in LPT 
particles continue to stream through. In full dark matter simulations particles stick 
together after orbit crossing, locked inside the high density regions called halos. 
In LPT the particles continue to stream out of these high density regions, effectively 
smearing these halos of to a few Mpc/h. As a result the power spectrum in any LPT is
below the linear power spectrum (at $z=0$). However, the seeds of halo formation 
have been imprinted already at the LPT level, and show up in the transfer functions, 
which essentially account for the artificial smearing of the high density regions by a 
smoothing radius that is independent of the position and hence can be represented 
as a $k$ dependent transfer function. Multiplying LPT power spectrum with the square 
of the transfer function $\tilde{T}(k)$ reverses this smearing. The halos, at least massive 
ones, exist in LPT, but their mass is spread out to such a large radius that they are 
not visible except for the largest ones. Reversing this with the transfer function brings 
the total halo mass to a smaller radius, and produces a better defined 1-halo term. 
These halos are still too diffuse relative to the N-body simulations, but contain all of 
the mass that has collapsed. As a consequence this term gives the correct 1-halo 
amplitude at low $k$ ($k \sim 0.2$h/Mpc for $z=0$). 

The transfer function $\tilde{T}(k)$ only goes so far in improving the relation between 
LPT and full dark matter N-body simulation. On smaller scales the dependence of the 
halo profile on the halo mass becomes important: halos have internal profiles, which 
depend on the halo mass, with larger halos having a larger extent: the relation can no 
longer be described in terms of a spatially independent transfer function. This is the 
stochasticity (sometimes called mode coupling) term $P_{J}(k)$. To improve the relation 
between LPT and N-body simulations one needs to further compress the halo profiles of 
the halo blobs as defined by the transfer function corrected LPT to the true halo profiles. 
The typical scale for this corresponds to the typical halo radius squared, averaged over 
the halo mass, which is of the order of 1Mpc/h, a smaller scale than the LPT smearing scale 
inside the transfer function. This process only redistributes the halo mass while conserving 
the total collapsed mass, so the process only becomes important at higher $k$ and does 
not affect clustering at $k < 0.2$h/Mpc. 

Note that both the transfer function correction term and the stochasticity term essentially try 
to do the same thing: correct the overly diffuse nature of halos in LPT to a more compact 
form found in N-body simulations. The only difference is that the transfer function part 
accounts for the halo mass (or spatial position) independent part, and dominates at low 
$k$, while stochasticity term depends on halo mass, and hence spatial position, and dominates 
at high $k$. They are also both determined by terms beyond PT, and become essentially 
free functions at high $k$. For the power spectrum there is therefore no obvious reason 
for splitting effects beyond PT into the two terms, and indeed in this paper we also defined 
$T(k)$ as a combined effect. Similarly, in \cite{2015PhRvD..91l3516S} both terms are put together 
into a single effective 1-halo $P_{\rm BB}(k)$ term, 
\be
P_{\rm dm}=P_{\rm 1LPT}+\big(T_{\rm 1LPT}^2-1\big)P_{\rm 1LPT}
=P_{\rm 1LPT}+\big(\tilde{T}_{\rm 1LPT}^2-1 \big)P_{\rm 1LPT}+P_{J} \equiv P_{\rm 1LPT}+P_{\rm BB},
\ee
where $P_{\rm BB}(k)$ is the non-perturbative part, that puts together all of the terms that 
cannot be computed in 1LPT, which in \cite{2014MNRAS.445.3382M} was modeled as 
$P_{\rm BB}(k)=F(k)A_0(1-R_{1h,2}^2k^2+R_{1h,4}^4k^4+...)$. 
Here $A_0$ is the amplitude of effective 1-halo term and $R_{1h,2i}$ are 
connected to the various moments of halo radius averaged over the halo profile and 
halo mass function, typically of the order of 1Mpc/h. For consistency of the halo model 
one needs to compensate the 1-halo term at very low $k$, and $F(k)$ is the compensation 
term, which in \cite{2015PhRvD..91l3516S} was modeled as $F(k)=1-1/(1+k^2R^2)$, 
where $R \sim 25$Mpc/h. This term is not associated with any physical scale, and is instead 
related to the 1LPT streaming error encoded in the leading EFT parameter (figure \ref{fig2}). 

\begin{figure*}[tb!]
\centering
\includegraphics[scale=0.53]{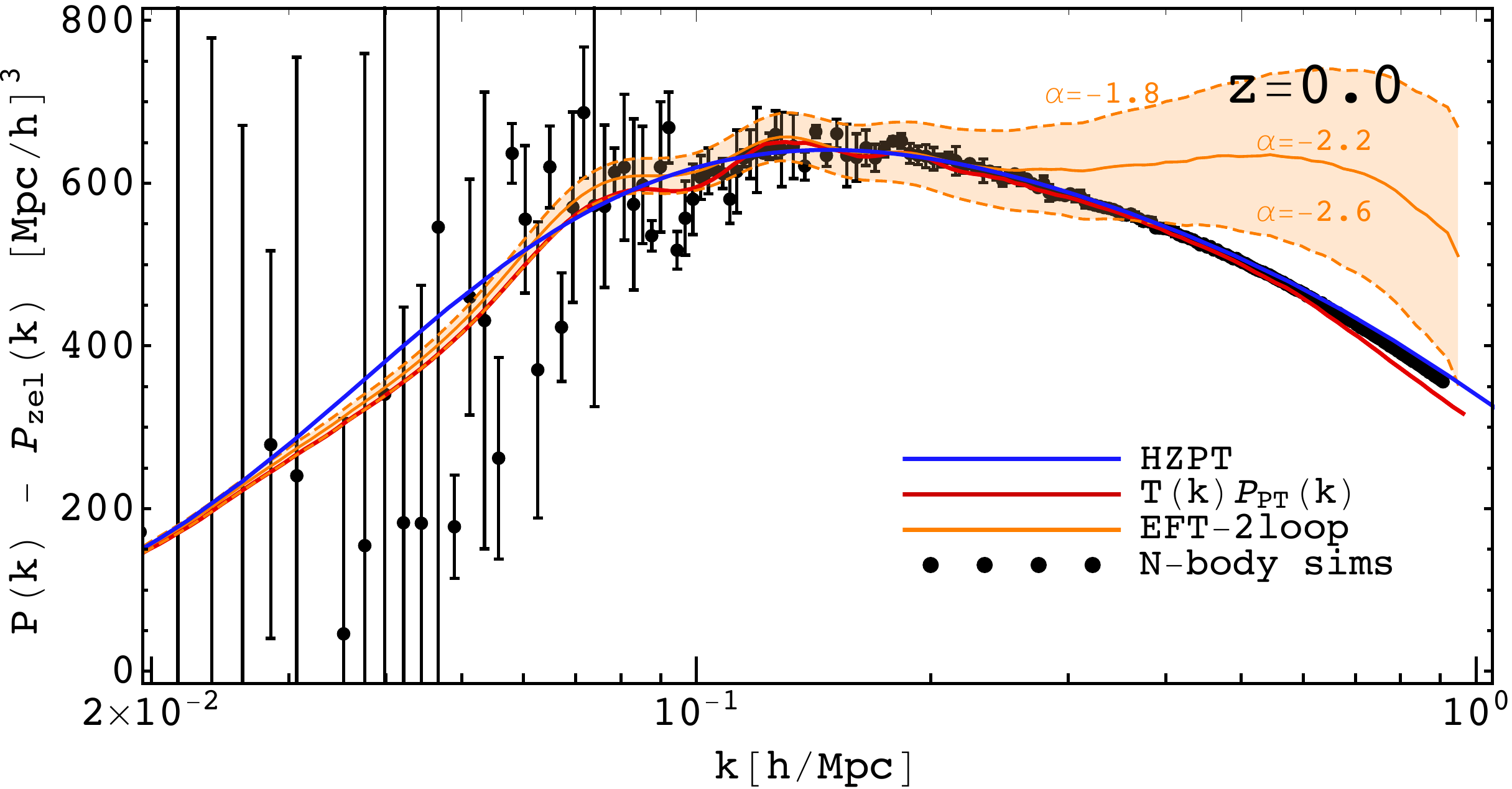}
\caption{
Difference of nonlinear dark matter power spectrum $P_{\rm dm}$ and $P_{\rm 1LPT}$ is shown 
for various models. We show N-body simulation results (as black points) and compare them to the 
transfer function approaches (red solid line). For comparison we show the HZPT model 
\cite{2015PhRvD..91l3516S} (blue solid line) and the two-loop EFT results from 
Eq. \eqref{eq:eft2loop} (orange solid line). For two-loop EFT results we added the (orange) band 
showing the results for $\alpha_{\rm SPT,1-loop} = -2.2 \pm 0.4~[h/{\rm Mpc}]^2$.
Results are shown for redshift $z=0.0$.
}
\label{fig7}
\end{figure*}

Since this compensation term is determined at low $k$, where stochasticity for 2LPT 
can be neglected, one can derive it from 2LPT transfer function. This is less true for 
1LPT, so the difference between 1LPT and 2LPT can be viewed as the source of 
stochasticity for 1LPT. The effect of compensation is typically below 1\%, so quite small. 
Since we have measured this term without sampling variance from 2LPT we can look 
at its scale dependence at low $k$, which is usually sampling variance dominated. 
In figure \ref{fig7} we plot this term against the $F(k)$ model of \cite{2015PhRvD..91l3516S}. 
We see that this term is only approximately modeled by the $F(k)$ ansatz above at very low $k$, 
although in practical terms this does not make much difference. 
In particular, the small EFT generated BAO effects discussed in previous section are not captured by 
$P_{\rm BB}(k)$ ansatz (which was also noted in \cite{2015PhRvD..91l3516S}). 
We note however that our results at low $k$ are only approximate, both because of 
stochasticity of 2LPT (at 10\% level) and because our analytic $P_{\rm 2LPT}(k)$ does 
not contain terms beyond 1-loop. The EFT term is similar to a 1-halo term, meaning 
that it is flat in $k$, around $k \sim 0.1-0.2$h/Mpc. This is because the linear power 
spectrum scales as $k^{-2}$ in that range. This is where one can merge the two 
approaches. In figure \ref{fig7} we show an attempt where we do this, defining maximum 
in $P_{\rm dm}-P_{\rm 1LPT}$ in EFT model to switch from EFT approach to halo model 
approach, which happens roughly at $k \sim 0.2$h/Mpc. EFT model of Eq. \eqref{eq:eft1loop} 
uses 1 EFT parameter. 

In \cite{2015PhRvD..91l3516S} both $A_0$ and $F(k)$ were extracted from SPT at low $k$ ($0.02{\rm h/Mpc}<k<0.03{\rm h/Mpc}$), 
under the assumption that the EFT correction is small. In figure \ref{fig7} we see that this 
correction is indeed very small there, but determining both parameters from a narrow range of $k$
induces degeneracies between them. A more robust approach would be to determine $A_0F(k)$
using simulations at higher $k$, or using EFT approach, where EFT parameters are fitted 
to simulations. However, since the effect of $F(k)$ is less than 1\%, in practice one is simply 
trading one free parameter, $A_0$, for another free parameter, that of EFT model. 
It may be possible to connect EFT and halo model approaches in the regime of overlap where 
both are applicable.
 
\section{Conclusions}
\label{conc}

The goal of the present analysis is to clarify some connections between the different 
PT schemes (LPT and SPT), the corresponding full dark matter solutions, and the halo 
formation process. In 1-D, 1LPT is exact at the PT level, corresponds to the SPT at 
infinite loop order, but does not give exact solution because of sheet crossings 
(``halo'' formation) that are beyond PT. 
One can parameterize the PT ignorance by a transfer function 
squared, defined as a ratio between true power spectrum and PT power spectrum, and 
one can expand this ignorance into a Taylor series which starts as $1+\alpha k^2$, 
which corresponds to the EFT expansion
at lower PT orders. At higher orders (beyond the two loop power spectrum) this 
equivalence does not hold any longer, and one would in principle need different transfer 
functions for different PT orders of the overdensity field.
In addition, one can decompose the ignorance of PT into a part that is correlated with the true 
density, and a part that is not, called stochasticity. 
The latter dominates on small scales, 
and correlated part dominates on large scales. The advantage of this split is that the 
correlated part can be used to predict higher order correlations of dark matter from 
the corresponding higher order correlations of the PT field, with no extra parameters 
as long as stochasticity can be ignored. In 1-d this EFT expansion has a larger radius of 
convergence than the SPT series, and knowing the two allows one to order the 
combined SPT+EFT expansion. In practice, we find 1 parameter EFT suffices for 
the range of 2-loop SPT. The improvements at 1-loop and 2-loop are impressive 
at low $k$, but get progressively harder as we push to higher $k$ where SPT 
series is very slowly convergent. Applying the EFT expansion to 1LPT, 
which corresponds to infinite loop SPT, is always better than the SPT+EFT expansion 
at the same order. There is only one EFT expansion, and only one set of EFT 
parameters, parameterizing the ignorance that can be applied to both SPT and 1LPT. 

In 3-d the situation is more complicated due to the lack of an exact PT solution. 
One has to compute higher order displacement fields, which can be computed in LPT, 
but their corresponding higher order LPT contributions become less and less reliable, 
and analytic expressions for the density field cannot be readily computed. It becomes more 
reliable to keep all the terms at a given PT order, which corresponds to $i$-loop SPT. 
For each LPT and SPT scheme there is a corresponding PT ignorance expansion 
in terms of EFT parameters that can be defined, and in general there is considerable running 
of EFT parameters as a function of $k$. For some schemes, such as 1-loop and 2-loop SPT, 
we expect there is little or no running at low $k$, and we indeed find that up to $k=0.1$h/Mpc 
for 1-loop and $k=0.2$h/Mpc for 2-loop. 
We also explored hybrid PT approaches, 
combining the full power spectrum of 1LPT with the 1-loop or 2-loop SPT terms missing from 
1LPT, but this does not extend the constancy of EFT parameter to a higher $k$ 
(and similarly for LEFT \cite{2015JCAP...09..014V}). All wiggle models include damping of the wiggles by 
IR modes, and we provide a simple derivation of the damping, including specific application to BAO. 

Finding a single EFT parameter is certainly desirable from the point of view of having a small 
number of free parameters, but does not address the issue of correctness of the EFT approach. 
To address that we produced N-body simulations with and without wiggles in the power spectrum to test 
these different EFT approaches. We ask the question whether the addition of EFT parameter 
improves the BAO wiggle description over the corresponding PT case without EFT. 
In all cases we find that the EFT term is able to reproduce BAO wiggles better than the 
corresponding PT models, supporting the basic idea of EFT. However, 
in many of the models PT alone does quite a good job in describing the BAO, and the residuals 
that EFT terms improve are very small. For $k>0.2$h/Mpc (at $z=0$) the difficulties of having 
reliable LPT or SPT higher order calculations, the running of EFT parameters, and the stochasticity 
all contribute to diminishing returns, and it appears difficult to extend EFT approach meaningfully 
beyond this $k$ without introducing additional free parameters. 
We also discuss the connection of EFT to the halo model \cite{2015PhRvD..91l3516S}, arguing that there is a regime 
around $k \sim 0.2$h/Mpc where both descriptions are valid, allowing one to connect the two. 

The concept of the transfer function describing the nonlinear 
effects beyond PT without generating stochasticity on large scales, is a useful tool that ensures 
there is only one such function that can be applied to correlations at all orders. This function must 
scale as $k^2$ at low $k$, where it is determined by a single number that can describe many different 
statistics. However, there is a fundamental difficulty in applying this concept to LSS, in that it works 
best at the lowest $k$. But in LSS these scales are of little interest to be modeled precisely beyond 
the linear theory, because nonlinear effects are small, and the sampling variance errors, which scale 
as the inverse square root of the number of modes, are large. The nonlinear effects are a few percent 
for $k<0.1$h/Mpc and we are unlikely to reach this level of precision observationally in the 
near future, and for $k>0.2$h/Mpc the EFT modeling with a single parameter breaks down at a sub-percent level, leaving 
0.1h/Mpc$<k<$0.2h/Mpc as the range where these models can have their main applicability. 
This is also the range where primordial wiggles still have some signatures in the nonlinear power spectrum. 
These statements all apply to $z=0$, and for higher redshifts the corresponding scale is shifted to higher $k$. 

\acknowledgments
We thank Tobias Baldauf, Leonardo Senatore, Martin White and Matias Zaldarriaga for useful disussions. 
We thank Matt McQuinn and Martin White for 1-d simulation and theory data. 
Z.V. is supported in part by the U.S. Department of Energy contract to SLAC no. DE-AC02-76SF00515.
U.S. is supported in part by the NASA ATP grant NNX12AG71G. 

\appendix

\section{Construction of the no-wiggle power spectra}
\label{app:nowiggle}

Here we summarize the methods of smoothing of the wiggles produced by baryon acoustic oscillations (BAO). We look at two 
filtering methods. The first employes the spherical Gaussian filters on spherical 
3D power spectra and the second employs a 1D Gaussian filter on the logarithmic scale. The latter 
turns out to be a superior method, recovering correctly small and large scale limits. 
Finally, we make use of the basis splines (B-spline) interpolation in order to achieve 
BAO smoothing, allowing us to impose additional constraints on the resultant power 
spectrum. This procedure enables us to construct no-wiggle spectra 
that have the same dispersions $\sigma_8^2$ and  $\sigma_v^2$ as the wiggle spectra. 
This is useful since we would want that our wiggle and no-wiggle power spectra 
exhibit same broadband nonlinear evolution. 
Additional advantages of the spline-based method are computational efficiency 
and automatization. 

\begin{figure}[t!]
    \centering
    \includegraphics[scale=0.72]{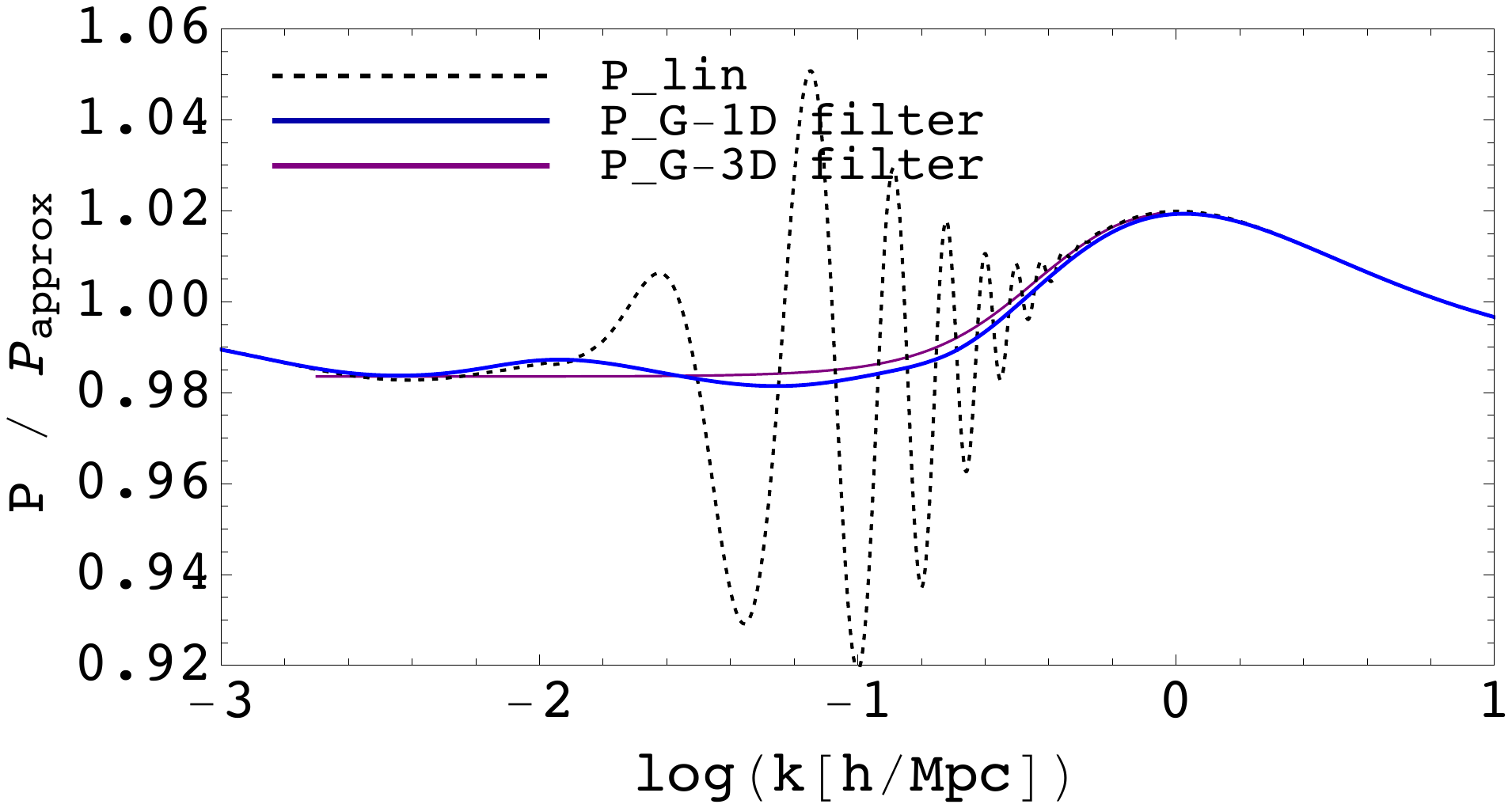}
    \caption{\small Linear power spectrum is shown in dotted back line. 
    Solid blue line is the smoothing result obtained by applying the 1D Gaussian filter, 
    and solid purple line is obtained applying 3D Gaussian filter in linear rather than 
    logarithmic $k$-spacing (note that for low $k$ the smooth power spectrum is underestimated). 
    All the lines are divided by the approximation of the no-wiggle power spectrum $P_{\rm approx}$ 
    given by \cite{1998ApJ...496..605E}.   
    }
    \label{fig:GaussFilter}
\end{figure}

We first explore the smoothing of the power spectra using Gaussian filtering. First we
note that it is useful to divide the initial power spectra by some approximate (wiggle free) 
curve \footnote{We use the approximation result given in \cite{1998ApJ...496..605E},
but alternative methods like BBKS \cite{1986ApJ...304...15B} would work equally well. 
Alternatively, one can use either some power law form $k^n$, with $n\propto1.5$, or 
actual output of  the Boltzmann codes with no baryons.}, in order to reduce the amplitude 
range of the broad band power. After smoothing is preformed, we retrieve the total 
no-wiggle power spectrum by multiplying the result back by this approximate curve. 
Final no-wiggle power spectrum is then given as
\begin{align}
P_{\rm nw} (k)= P_{\rm approx}(k) \mathcal{F}\left[ P(k)/P_{\rm approx}(k)\right],
\end{align}
where $P(k)$ is the initial power spectrum that needs to be smoothed,
and $P_{\rm approx}$ is the initial broad band approximation curve (given by e.g.
 \cite{1998ApJ...496..605E,1986ApJ...304...15B}). 

First, we explore a 3d Gaussian filter in linear $k$ spacing:
\begin{align}
\mathcal{F}_G(\vec {k})=\frac{1}{\sqrt{(2\pi)^3}\lambda^3}
\exp\left(-\frac{1}{2\lambda^2}|\vec {k}|^2\right),
\end{align}
where $\lambda$ typically takes a value around 0.05 Mpc$/h$.  
For the no-wiggle power spectra we then have
\begin{align}
P_{\rm nw} (k) &= \int d^3q~P(q) \mathcal{F}_G(|\vec {k}-\vec {q}|) \nonumber\\
                      &= \frac{\sqrt{2}}{\sqrt{\pi} \lambda} \int dq~q^2 P(q) 
                      \exp\left(-\frac{1}{2\lambda^2}(q^2+k^2)\right) \frac{\sinh\left(kq/\lambda^2\right)}{kq}.
\end{align}
The problem that emerges when implementing this method is at the low $k$ limit of the spectra. 
Since the smoothing radius is about the size of wiggle wavelength and since the first wiggle 
starts at the distance closer to the $k=0$ than the typical wiggle size, smoothing introduces a mismatch 
of power amplitude at the $k=0$ limit (see figure \ref{fig:GaussFilter}). 
Note also that allowing $\lambda$ to vary with scale, $\lambda \to \lambda(k)$, so that 
$\lambda \to 0$ as $k\rightarrow0$, is also not an optimal solution since this procedure could 
itself introduce some spurious wiggle-like features, and thus requires more complicated filter ansatz.

\begin{figure}[t!]
    \centering
    \includegraphics[scale=0.47]{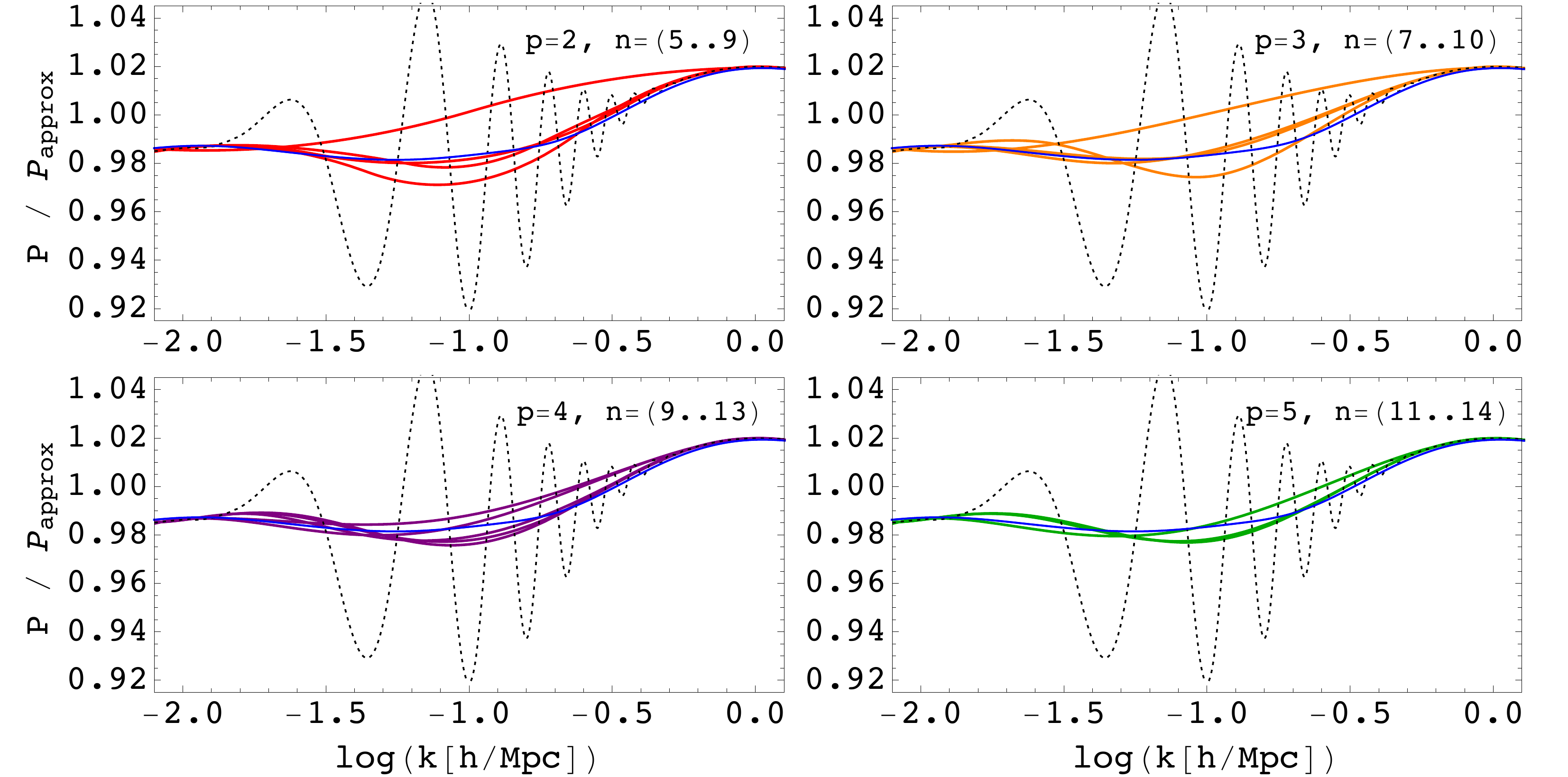}
    \caption{\small Grid of B-spline smoothing curves (colored solid lines), varying 
    in spline degree $p$ and the number of knots $n$. Linear power spectrum is shown 
    in dotted back line. For comparison smoothing result obtained from applying the 
    1D Gaussian filter is shown as a solid blue line. 
    All the lines are divided by the approximation no-wiggle power spectrum $P_{\rm approx}$ 
    given by \cite{1998ApJ...496..605E}.   
    }
    \label{fig:BSplineFamily}
\end{figure}

Transforming the power spectra to the logarithmic variable, so that the range 
covers the whole real axes, enables us to use simple 1D filters. 
In this case, we get a simple expression 
\begin{align}
P_{\rm nw} \left(10^{k_{\log}}\right) &= \frac{1}{\sqrt{2\pi}\lambda}\int d q_{\log}
~P\left(10^{q_{\log}} \right) \exp \left(-\frac{1}{2\lambda^2}(k_{\log}-q_{\log})^2\right),
\end{align} 
where $k_{\log}$ and $q_{\log}$ are variables that take values from the whole real axis, 
and is $\lambda$ a parameter with typical value $~0.25h/$Mpc. 
Note that since wiggles are now not equidistantly spaced (being more compressed 
at higher $k$), it is beneficial to introduce a slight scale dependence of $\lambda$, 
increasing the value towards higher $k$. Result of applying these filters on the linear 
power spectrum is also shown in figure \ref{fig:GaussFilter}. We see that at the low $k$ 
smooth power spectrum is underestimated if one uses the 3D filters in linear spacing.

Alternative to using the explicit filters to smooth the BAO wiggles is using the basis splines 
(B-splines)\footnote{see e.g. http://mathworld.wolfram.com/B-Spline.html}. 
B-splines offer a convenient way to approximate our wiggle curves. 
A B-spline is a generalisation of the B\' ezier curve. We can define a knot vector 
$\textbf{T}=\left[ t_0,~t_1,\ldots,~t_m\right]$, where $\textbf{T}$ is a nondecreasing 
sequence with $t_i$ in $\left[0,1\right]$, and the set of control points $C_0,~\ldots,~C_n$. 
Degree of the B-spline is then given as $p=m-n-1$ and the knots $t_{p+1},~\ldots,~t_{m-p-1}$ 
are called internal knots. The basis functions are defined as
\begin{align}
N_{i,0} (t) & = \begin{cases} 1 &\mbox{if } t_i \leq t < t_{i+1} \\ 0 & \mbox{otherwise } \end{cases}, \nonumber\\
N_{i,j} (t) & = \frac{t-t_i}{t_{i+j}-t_i}N_{i,j-1}(t)+\frac{t_{i+j+1}-t}{t_{i+j+1}-t_{i+1}}N_{i+1,j-1}(t),
\end{align}
where $j=1,~2,~\ldots,~p$. We can define the B-spline power spectrum approximation as
\begin{align}
P_{n,p}(k) = \sum_{i=0}^n C_i N_{i,p}(k).
\end{align}
Knot values determine the extent of the control of control points.
In order to achieve smoothing $C_i$ can be determined by the some regression methods 
and we use a simple linear model fit.

\begin{figure}[t!]
    \centering
    \includegraphics[scale=0.445, trim=1.6cm 0 2.5cm 0, clip]{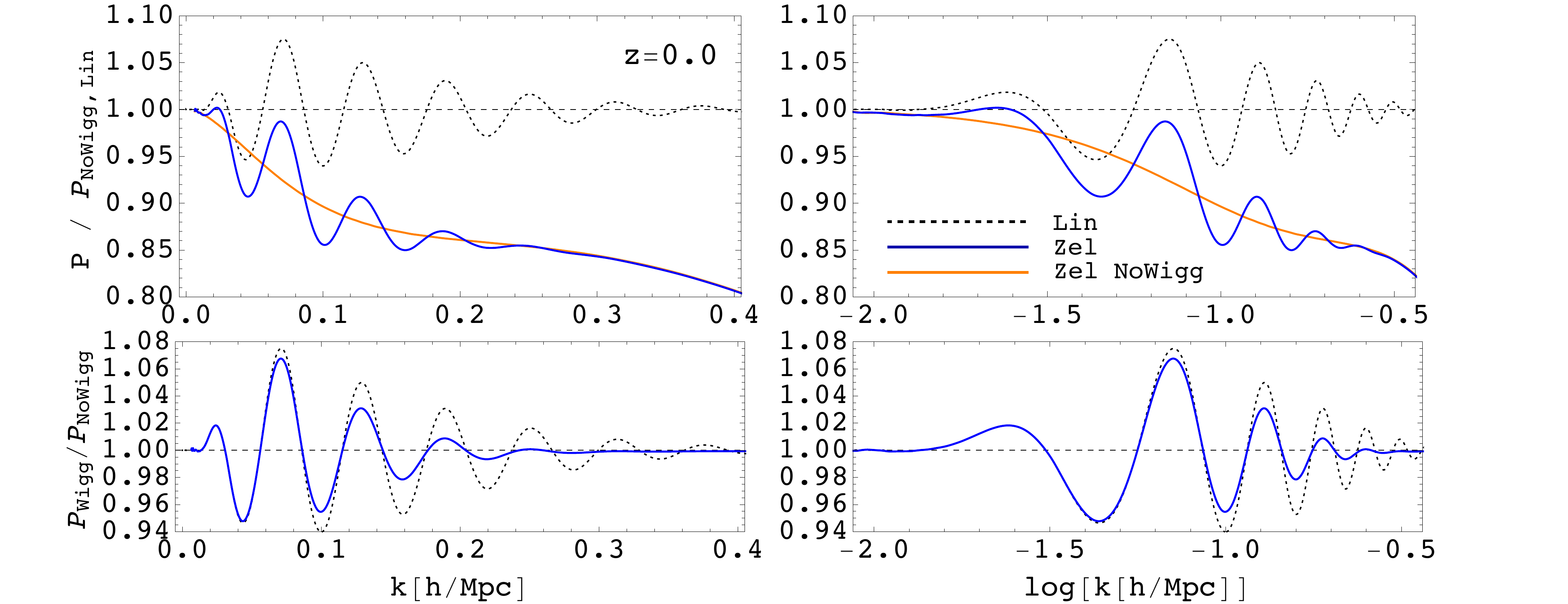}
    \caption{\small In upper panels the Zel'dovich power spectrum is shown as the blue line, and
    the resulting smoothed version as the orange line. Smoothed version is obtained by computing the 
    Zel'dovich power spectrum using the smoothed linear spectrum form Eq. \eqref{eq:avvPS}. 
    Dotted line is linear power spectrum, and all the lines are divided by the smooth version of linear 
    spectrum given by Eq. \eqref{eq:avvPS}. In lower panels the Zel'dovich power spectrum is shown divided 
    by the smooth version (solid orange line). For the reference, the linear power spectrum is also shown 
    divided by the smooth version of itself (black dotted line). Left and right panels differ in $k-$spacing: 
    linear vs. logarithmic, in order to stress the agreement on both large and small scales.
    }
    \label{fig:SmoothZel}
\end{figure}

One can construct a family of curves that approximate the original wiggle spectrum. 
The family of such curves is shown in figure \ref{fig:BSplineFamily} for spline degrees $p=2\ldots5$ 
with a various number of knots. This family allows us now to impose certain constraints on the 
final no-wiggle spectra $P_{\rm nw}$. 
One such constraint can be velocity dispersion
\begin{align}
\sigma^2_v=\frac{1}{3}\int \frac{d^3q}{(2\pi)^3}\frac{P(k)}{q^2}.
\end{align}
A similar constraint is the $\sigma_8$ defined in Eq. \eqref{eq:sigv_sig8}. 
Note that we can impose several conditions at the 
same time as long the constructed B-spline curves are distributed widely enough. 
We can thus construct the final smooth spectra as a weighted average over the spline curves
\begin{align}
P_{\rm nw}(k)=\frac{1}{N}\sum_{s=1}^N w_s P_s (k),
\label{eq:avvPS}
\end{align}
where summation in $s$ includes all the eligible approximations given by the knot 
vector $\textbf{T}$ and degree of the B-splines $p$. Imposing the integral 
constraints like $\sigma^2_v$ and $\sigma^2_8$ on the $P_{\rm nw}$ places constraints 
on the weights $w_i$. 

As a simple example of the smoothing procedure, we can have a look a the Zel'dovich 
power spectrum. In figure \ref{fig:SmoothZel} we show the results of the 
Zel'dovich power spectrum computed using the power spectrum with and without the wiggles. 
Note that in this case it is very useful that the smoothed linear spectrum satisfies the integral 
constraints and have the same $\sigma_8$ and $\sigma_v$ values. 
These ensures that on the large as well as small scales away from the 
BAO region two spectra will agree precisely.

\bibliographystyle{JHEP}
\bibliography{cosmo,cosmo_preprints}

\providecommand{\href}[2]{#2}\begingroup\raggedright\begin{thebibliography}{10}

\bibitem{2012JCAP...07..051B}
D.~{Baumann}, A.~{Nicolis}, L.~{Senatore}, and M.~{Zaldarriaga}, {\it
  {Cosmological non-linearities as an effective fluid}},  {\em \jcap} {\bf 7}
  (July, 2012) 51, [\href{http://xxx.lanl.gov/abs/1004.2488}{{\tt
  arXiv:1004.2488}}].

\bibitem{2012JHEP...09..082C}
J.~J.~M. {Carrasco}, M.~P. {Hertzberg}, and L.~{Senatore}, {\it {The effective
  field theory of cosmological large scale structures}},  {\em Journal of High
  Energy Physics} {\bf 9} (Sept., 2012) 82,
  [\href{http://xxx.lanl.gov/abs/1206.2926}{{\tt arXiv:1206.2926}}].

\bibitem{2013JCAP...08..037P}
E.~{Pajer} and M.~{Zaldarriaga}, {\it {On the renormalization of the effective
  field theory of large scale structures}},  {\em \jcap} {\bf 8} (Aug., 2013)
  37, [\href{http://xxx.lanl.gov/abs/1301.7182}{{\tt arXiv:1301.7182}}].

\bibitem{2015JCAP...02..013S}
L.~{Senatore} and M.~{Zaldarriaga}, {\it {The IR-resummed Effective Field
  Theory of Large Scale Structures}},  {\em \jcap} {\bf 2} (Feb., 2015) 13,
  [\href{http://xxx.lanl.gov/abs/1404.5954}{{\tt arXiv:1404.5954}}].

\bibitem{2014JCAP...05..022P}
R.~A. {Porto}, L.~{Senatore}, and M.~{Zaldarriaga}, {\it {The Lagrangian-space
  Effective Field Theory of large scale structures}},  {\em \jcap} {\bf 5}
  (May, 2014) 22, [\href{http://xxx.lanl.gov/abs/1311.2168}{{\tt
  arXiv:1311.2168}}].

\bibitem{2014JCAP...07..057C}
J.~J.~M. {Carrasco}, S.~{Foreman}, D.~{Green}, and L.~{Senatore}, {\it {The
  Effective Field Theory of Large Scale Structures at two loops}},  {\em \jcap}
  {\bf 7} (July, 2014) 57, [\href{http://xxx.lanl.gov/abs/1310.0464}{{\tt
  arXiv:1310.0464}}].

\bibitem{2015arXiv150705326F}
S.~{Foreman}, H.~{Perrier}, and L.~{Senatore}, {\it {Precision Comparison of
  the Power Spectrum in the EFTofLSS with Simulations}},  {\em ArXiv e-prints}
  (July, 2015) [\href{http://xxx.lanl.gov/abs/1507.0532}{{\tt
  arXiv:1507.0532}}].

\bibitem{2002PhR...367....1B}
F.~{Bernardeau}, S.~{Colombi}, E.~{Gaztanaga}, and R.~{Scoccimarro}, {\it
  {Large-scale structure of the Universe and cosmological perturbation
  theory.}},  {\em \physrep} {\bf 367} (2002) 1--128.

\bibitem{2013MNRAS.429.1674C}
J.~{Carlson}, B.~{Reid}, and M.~{White}, {\it {Convolution Lagrangian
  perturbation theory for biased tracers}},  {\em \mnras} {\bf 429} (Feb.,
  2013) 1674--1685, [\href{http://xxx.lanl.gov/abs/1209.0780}{{\tt
  arXiv:1209.0780}}].

\bibitem{2014JCAP...01..010B}
D.~{Blas}, M.~{Garny}, and T.~{Konstandin}, {\it {Cosmological perturbation
  theory at three-loop order}},  {\em \jcap} {\bf 1} (Jan., 2014) 10,
  [\href{http://xxx.lanl.gov/abs/1309.3308}{{\tt arXiv:1309.3308}}].

\bibitem{1980lssu.book.....P}
P.~J.~E. {Peebles}, {\em {The large-scale structure of the universe}}.
\newblock Research supported by the National Science Foundation.~Princeton,
  N.J., Princeton University Press, 1980.~435 p., 1980.

\bibitem{2000MNRAS.318..203S}
U.~{Seljak}, {\it {Analytic model for galaxy and dark matter clustering}},
  {\em \mnras} {\bf 318} (Oct., 2000) 203--213,
  [\href{http://xxx.lanl.gov/abs/astro-ph/0001493}{{\tt astro-ph/0001493}}].

\bibitem{2001ApJ...546...20S}
R.~{Scoccimarro}, R.~K. {Sheth}, L.~{Hui}, and B.~{Jain}, {\it How many
  galaxies fit in a halo? constraints on galaxy formation efficiency from
  spatial clustering},  {\em \apj} {\bf 546} (Jan., 2001) 20--34.

\bibitem{2002PhR...372....1C}
A.~{Cooray} and R.~{Sheth}, {\it {Halo models of large scale structure}},  {\em
  \physrep} {\bf 372} (Dec., 2002) 1--129,
  [\href{http://xxx.lanl.gov/abs/astro-ph/0206508}{{\tt astro-ph/0206508}}].

\bibitem{2011A&A...527A..87V}
P.~{Valageas} and T.~{Nishimichi}, {\it {Combining perturbation theories with
  halo models}},  {\em \aap} {\bf 527} (Mar., 2011) A87,
  [\href{http://xxx.lanl.gov/abs/1009.0597}{{\tt arXiv:1009.0597}}].

\bibitem{2014MNRAS.445.3382M}
I.~{Mohammed} and U.~{Seljak}, {\it {Analytic model for the matter power
  spectrum, its covariance matrix and baryonic effects}},  {\em \mnras} {\bf
  445} (Dec., 2014) 3382--3400, [\href{http://xxx.lanl.gov/abs/1407.0060}{{\tt
  arXiv:1407.0060}}].

\bibitem{2012JCAP12011T}
S.~{Tassev} and M.~{Zaldarriaga}, {\it {Estimating CDM particle trajectories in
  the mildly non-linear regime of structure formation. Implications for the
  density field in real and redshift space}},  {\em \jcap} {\bf 12} (Dec.,
  2012) 11, [\href{http://xxx.lanl.gov/abs/1203.5785}{{\tt arXiv:1203.5785}}].

\bibitem{2015arXiv150702255B}
T.~{Baldauf}, E.~{Schaan}, and M.~{Zaldarriaga}, {\it {On the reach of
  perturbative methods for dark matter density fields}},  {\em ArXiv e-prints}
  (July, 2015) [\href{http://xxx.lanl.gov/abs/1507.0225}{{\tt
  arXiv:1507.0225}}].

\bibitem{2015arXiv150706665B}
D.~{Blas}, S.~{Floerchinger}, M.~{Garny}, N.~{Tetradis}, and U.~A. {Wiedemann},
  {\it {Large scale structure from viscous dark matter}},  {\em ArXiv e-prints}
  (July, 2015) [\href{http://xxx.lanl.gov/abs/1507.0666}{{\tt
  arXiv:1507.0666}}].

\bibitem{2015PhRvD..91b3508V}
Z.~{Vlah}, U.~{Seljak}, and T.~{Baldauf}, {\it {Lagrangian perturbation theory
  at one loop order: Successes, failures, and improvements}},  {\em \prd} {\bf
  91} (Jan., 2015) 023508, [\href{http://xxx.lanl.gov/abs/1410.1617}{{\tt
  arXiv:1410.1617}}].

\bibitem{2009JCAP...10..007M}
P.~{McDonald} and U.~{Seljak}, {\it {How to evade the sample variance limit on
  measurements of redshift-space distortions}},  {\em Journal of Cosmology and
  Astro-Particle Physics} {\bf 10} (Oct., 2009) 7--+,
  [\href{http://xxx.lanl.gov/abs/0810.0323}{{\tt arXiv:0810.0323}}].

\bibitem{2015arXiv150207389M}
M.~{McQuinn} and M.~{White}, {\it {Cosmological perturbation theory in 1+1
  dimensions}},  {\em ArXiv e-prints} (Feb., 2015)
  [\href{http://xxx.lanl.gov/abs/1502.0738}{{\tt arXiv:1502.0738}}].

\bibitem{2015arXiv150507098B}
T.~{Baldauf}, E.~{Schaan}, and M.~{Zaldarriaga}, {\it {On the reach of
  perturbative descriptions for dark matter displacement fields}},  {\em ArXiv
  e-prints} (May, 2015) [\href{http://xxx.lanl.gov/abs/1505.0709}{{\tt
  arXiv:1505.0709}}].

\bibitem{Feng...inprep}
Y.~{Feng}, M.~{Chu}, and U.~{Seljak}, {\it {Mocking Large Scale Structure with
  Fast Particle-Mesh (fastPM)}},  \href{http://xxx.lanl.gov/abs/in prep.}{{\tt
  in prep.}}

\bibitem{2015JCAP...09..014V}
Z.~{Vlah}, M.~{White}, and A.~{Aviles}, {\it {A Lagrangian effective field
  theory}},  {\em \jcap} {\bf 9} (Sept., 2015) 14,
  [\href{http://xxx.lanl.gov/abs/1506.0526}{{\tt arXiv:1506.0526}}].

\bibitem{2015PhRvD..92d3514B}
T.~{Baldauf}, M.~{Mirbabayi}, M.~{Simonovi{\'c}}, and M.~{Zaldarriaga}, {\it
  {Equivalence principle and the baryon acoustic peak}},  {\em \prd} {\bf 92}
  (Aug., 2015) 043514, [\href{http://xxx.lanl.gov/abs/1504.0436}{{\tt
  arXiv:1504.0436}}].

\bibitem{2014JCAP...06..008T}
S.~{Tassev}, {\it {Lagrangian or Eulerian; real or Fourier? Not all approaches
  to large-scale structure are created equal}},  {\em \jcap} {\bf 6} (June,
  2014) 8, [\href{http://xxx.lanl.gov/abs/1311.4884}{{\tt arXiv:1311.4884}}].

\bibitem{2008PhRvD..77f3530M}
T.~{Matsubara}, {\it {Resumming cosmological perturbations via the Lagrangian
  picture: One-loop results in real space and in redshift space}},  {\em \prd}
  {\bf 77} (Mar., 2008) 063530, [\href{http://xxx.lanl.gov/abs/0711.2521}{{\tt
  arXiv:0711.2521}}].

\bibitem{2010JCAP...06..009F}
R.~{Flauger}, L.~{McAllister}, E.~{Pajer}, A.~{Westphal}, and G.~{Xu}, {\it
  {Oscillations in the CMB from axion monodromy inflation}},  {\em \jcap} {\bf
  6} (June, 2010) 9, [\href{http://xxx.lanl.gov/abs/0907.2916}{{\tt
  arXiv:0907.2916}}].

\bibitem{2014arXiv1406.0548M}
P.~D. {Meerburg}, D.~N. {Spergel}, and B.~D. {Wandelt}, {\it {Searching for
  oscillations in the primordial power spectrum}},  {\em ArXiv e-prints} (June,
  2014) [\href{http://xxx.lanl.gov/abs/1406.0548}{{\tt arXiv:1406.0548}}].

\bibitem{2015JCAP...05..007B}
T.~{Baldauf}, L.~{Mercolli}, M.~{Mirbabayi}, and E.~{Pajer}, {\it {The
  bispectrum in the Effective Field Theory of Large Scale Structure}},  {\em
  \jcap} {\bf 5} (May, 2015) 7, [\href{http://xxx.lanl.gov/abs/1406.4135}{{\tt
  arXiv:1406.4135}}].

\bibitem{2014arXiv1406.4143A}
R.~E. {Angulo}, S.~{Foreman}, M.~{Schmittfull}, and L.~{Senatore}, {\it {The
  One-Loop Matter Bispectrum in the Effective Field Theory of Large Scale
  Structures}},  {\em ArXiv e-prints} (June, 2014)
  [\href{http://xxx.lanl.gov/abs/1406.4143}{{\tt arXiv:1406.4143}}].

\bibitem{2015PhRvD..91l3516S}
U.~{Seljak} and Z.~{Vlah}, {\it {Halo Zel'dovich model and perturbation theory:
  Dark matter power spectrum and correlation function}},  {\em \prd} {\bf 91}
  (June, 2015) 123516, [\href{http://xxx.lanl.gov/abs/1501.0751}{{\tt
  arXiv:1501.0751}}].

\bibitem{1998ApJ...496..605E}
D.~J. {Eisenstein} and W.~{Hu}, {\it {Baryonic Features in the Matter Transfer
  Function}},  {\em \apj} {\bf 496} (Mar., 1998) 605.

\bibitem{1986ApJ...304...15B}
J.~M. {Bardeen}, J.~R. {Bond}, N.~{Kaiser}, and A.~S. {Szalay}, {\it {The
  statistics of peaks of Gaussian random fields}},  {\em \apj} {\bf 304} (May,
  1986) 15--61.

\end{thebibliography}\endgroup
\end{document}